\documentclass[aps,prd,twocolumn,groupedaddress,amssymb,eqsecnum,showpacs,
nofootinbib]{revtex4}
\usepackage{graphicx}
\usepackage{mathptm}
\usepackage{bm}
\usepackage{times}

\begin{document}

\preprint{UMHEP-461,CMU-HEP-05-07}

\title{An effective theory of initial conditions in inflation}

\author{Hael Collins}
\email{hael@physics.umass.edu}
\affiliation{Department of Physics, University of Massachusetts, 
Amherst MA\ \ 01003}
\author{R.~Holman}
\email{rh4a@andrew.cmu.edu}
\affiliation{Department of Physics, Carnegie Mellon University, 
Pittsburgh PA\ \ 15213}

\date{\today}

\begin{abstract}
We examine the renormalization of an effective theory description of a general initial state set in an isotropically expanding space-time, which is done to understand how to include the effects of new physics in the calculation of the cosmic microwave background power spectrum.  The divergences that arise in a perturbative treatment of the theory are of two forms:  those associated with the properties of a field propagating through the bulk of space-time, which are unaffected by the choice of the initial state, and those that result from summing over the short-distance structure of the initial state.  We show that the former have the same renormalization and produce the same subsequent scale dependence as for the standard vacuum state, while the latter correspond to divergences that are localized at precisely the initial time hypersurface on which the state is defined.  This class of divergences is therefore renormalized by adding initial-boundary counterterms, which render all of the perturbative corrections small and finite.  Initial states that approach the standard vacuum at short distances require, at worst, relevant or marginal boundary counterterms.  States that differ from the vacuum at distances below that at which any new, potentially trans-Planckian, physics becomes important are renormalized with irrelevant boundary counterterms.
\end{abstract}


\maketitle

\section{Introduction}
\label{intro}

This article is the second in a series that develops a renormalizable effective theory description of the initial state for inflation.  The first article \cite{greens} described the construction of an effective initial state in Minkowski space while this one generalizes the setting to an isotropically expanding space-time.  The main idea is to establish a perturbative description of the signals of new physics as it affects the short-distance structure of the state, showing explicitly how any divergences associated with this short-distance structure are renormalized.

The accelerated expansion of the universe during inflation \cite{textbooks} produces an extremely rapid growth in the size of a causally connected region while at the same time the Hubble size remains essentially unchanged.  The exact amount of expansion depends upon the details of the inflationary model, but with 60--70 $e$-folds of inflation, the entire universe seen today could have grown from a single, suitably small, causally connected region.  This mechanism can explain the extreme uniformity of the universe observed on large scales or at early times, but even more importantly, inflation also predicts a tiny departure from perfect homogeneity caused by quantum fluctuations which are similarly stretched to vast scales.  The spectrum of these fluctuations is exactly the form of the synchronized acoustic oscillations which have been measured precisely by the Wilkinson Microwave Anisotropy Probe (WMAP) \cite{wmap}.  

Most models for inflation have no difficulty in producing the required amount of inflation.  Typically they produce substantially more and as a consequence the structures on the largest cosmological scales today would have had their origin in quantum fluctuations which occurred at sub-Planckian lengths during inflation.  This peculiar feature of inflation, its ``trans-Planckian problem'' \cite{brandenberger}, suggests the possibility that extremely short-distance physics, well beyond that currently accessible in accelerator experiments, could be imprinted on the very largest of observable scales.  Although the term suggests physics above the Planck scale, here we shall generally refer to {\it any\/} sort of new physics sufficiently above the Hubble scale during inflation as ``trans-Planckian.''

With this opportunity and with the prospect of significantly better measurements of the cosmic microwave background and the large scale structure, it is increasingly important to have an accurate estimate of the generic trans-Planckian signal.  One approach for determining this signal is to choose a specific model for what happens above the Planck scale and then to calculate its corrections to the primordial fluctuation spectrum \cite{gary,transplanck,cliff,fate,ekp1}.  Such models have been very useful in providing an estimate of the typical size for the trans-Planckian signal. However, the details of a particular model may have relatively little motivation from lower energy phenomenology or might simply not correspond to what actually occurs in nature.

A second approach does not attempt to form a complete picture of the physics above the Planck scale, but rather seeks to develop an effective theory description of its signal \cite{greens,kaloper,schalm,schalm2,ekp2,emil}.  Any new physics near the Planck scale first appears as a generic set of effects during inflation which are specified by a small number of parameters.  While they assume specific values in any particular model for the new physics, to a low energy observer they simply appear as free parameters to be fixed experimentally.  The effective approach is based on a perturbative expansion that uses the smallness of the ratio of the two natural scales---$H$, the Hubble scale during inflation, and $M$, the scale associated with the possible new physics.  If the signal is suppressed only by $H/M$, it could well be observable either by future cosmic microwave satellites such as Planck \cite{planck} or, still more likely, by future surveys \cite{ska} of the large scale structure, which traces the same primordial fluctuations.

From the perspective of the effective theory principle, new physics can appear in either the time evolution of the inflaton and its fluctuations or in their ``initial'' states.  The first of these---how the system evolves---is more familiar since the evolution of the quantum fluctuations of the inflaton is determined by its interaction Hamiltonian.  The form of the general set of possible corrections that we can add to this Hamiltonian, encoding the effects of the unknown physics, is rather constrained by the space-time symmetries.  Given these constraints, the size of the corrections from the unknown physics relative to the leading prediction for an inflationary model is usually suppressed by a factor of $(H/M)^2$ \cite{kaloper}.  The other ingredient---the state of the inflaton---is more directly related to the trans-Planckian problem because it is the details of the initial state we have chosen which are being stretched to vast scales.  The leading correction from these effects is typically much less suppressed, scaling instead as $H/M$ \cite{gary,transplanck,cliff,fate,ekp1}.

The standard view is that the correct state to choose is the vacuum state.  In an expanding background, this vacuum corresponds to the maximally symmetric state that matches with the Minkowski space vacuum at very short distances, where the curvature of the space-time should be negligible.  This requirement is made so that when a quantum field theory is placed in a curved background, it inherits the same renormalizability it had in Minkowski space, although its behavior can be quite different on large scales.

Despite the apparent simplicity of this view, it contains an inherent ambiguity.  Even in Minkowski space, there is no reason to trust that the vacuum, defined in terms of the eigenstates of a low energy effective theory, has the correct structure when extrapolated to arbitrarily short distances.  The true vacuum, determined by the eigenstates of the complete theory, most likely departs from a perfect agreement with the extrapolated low energy vacuum on sufficiently small scales.  An effective theory description of an initial state then provides a method for characterizing these departures such that they only have a small effect on long-distance measurements and moreover do not lead to any uncontrolled divergences which would prevent a perturbative calculation \cite{greens}.

The effective theory principle \cite{eft} provides a powerful prescription for understanding some phenomena over a limited range of length or energy scales, while at the same time parameterizing the possible leading signals of unknown physics that lies just beyond those scales.  For a quantum field theory propagating through the bulk of space-time, this idea is implemented by first identifying all of the observed fields and symmetries of a physical system and then constructing all possible operators out of these fields that are invariant under all of the symmetries.  A finite, and usually small, subset of these operators describes the theory well at very large distances, or equivalently at low energies.  The remaining infinite set of operators represents the generic signal of the hidden degrees of freedom whose dynamics are associated with a much smaller characteristic length, $1/M$.  What makes the theory calculable at low energies, $E\ll M$, is that all of the operators in this second set are naturally suppressed by some $n^{\rm th}$ power of $E/M$, with only a finite number appearing at each order $n$.  For an experiment conducted at an energy $E$ and which measures an observable to an accuracy $\delta$, we need only include the set of operators whose order $n$ satisfies 
\begin{equation}
\left( {E\over M} \right)^n \ge \delta . 
\label{eftbnd}
\end{equation}
When an experiment finally probes energies of the order $E\sim M$, where the effective theory becomes nonpredictive, it uncovers the hidden degrees of freedom and their symmetries and the original effective theory is replaced by a new effective theory including these new fields which is applicable up to some still higher energy.

The point of an effective description of an initial state is similarly to divide the aspects of a state into components which are important either at long or at short distances.  A state constructed only from the former agrees with the standard vacuum at arbitrarily short distances.  Yet even for such states, if the difference between the actual state and the standard vacuum diminishes sufficiently slowly, some new divergences appear in the perturbative corrections.  These divergences only occur at the initial time, precisely where the state was defined, and so must be cancelled by adding new counterterms, localized at the initial boundary, which are marginal or relevant according to a boundary action.  

The short-distance components, which contain the effects of trans-Planckian physics, describe states which diverge from the standard vacuum at distances below $1/M$.  This behavior is completely consistent as an effective theory and only produces initial-time divergences which are cancelled by adding irrelevant counterterms to the boundary action.  As with the standard setting for an effective theory, for a measurement at low energies, which for inflation corresponds to the Hubble scale $H$, the effects of the trans-Planckian components of the state are suppressed by powers of $H/M$ leading to a completely predictive theory.

An effective theory of any form is always inherently applicable only up to a particular energy scale.  In an expanding background, this property also sets a limit to the earliest possible time at which it can be applied while still remaining perturbative \cite{schalm2,ekp2}.  While this limit implies that we should choose our ``initial'' state no earlier than that time at which redshifting would undo the suppression of the ratio $H/M$, it also implies that we should not use a formalism, such as the $S$-matrix, which would require integrating over any times earlier than this initial time.

What underlies the success of an overall space-time view such as the $S$-matrix in Minkowski space is that scales do not alter over time.  If we define some departure between the standard vacuum and the true vacuum at some very small scale in an asymptotic past, that scale remains unchanged during the period when the parts of the system interact and further on into some asymptotic future.  In contrast, with the continuous stretching of scales during inflation, it is not possible to apply a such an overall space-time view since the asymptotic states would need to be defined in a regime where all the features responsible for the shape of the microwave background would have been infinitesimally smaller than the Planck scale.  Instead, the correct approach is to use the Hamiltonian to evolve the entire system continuously forward, starting from a state defined at an appropriate initial time \cite{witten,schwinger,keldysh,kt,weinberginin}. 

The next section begins by describing how to set the initial state in a Robertston-Walker space-time by using a boundary condition specified along a spacelike surface.  To renormalize the theory, it is important to extract the exact behavior at asymptotically short distances, which is accomplished here by applying an adiabatic expansion for the state.  For a full time-dependent description of the effects of a general initial state, the theory must describe how the information in this initial state propagates forward, so in Sec.~\ref{prop} we construct a propagator which is also consistent with the initial boundary condition.

Our ultimate interest is to calculate the generic trans-Planckian signal in the microwave background \cite{twopoint} which, although a tree-level calculation, implicitly assumes that perturbative corrections are small and finite.  It is therefore necessary to establish the renormalizability of the the theory for a general initial condition.  This calculation begins in Sec.~\ref{renormalization} with a statement of an appropriate renormalization condition for this setting and its implications for the renormalization and running of the bulk parameters of the theory are presented in Sec.~\ref{bulk}.  The renormalization and running of the initial condition are examined in Sec.~\ref{boundary}, which shows how the renormalizable and nonrenormalizable classes of initial conditions are associated with relevant or irrelevant boundary counterterms respectively.  We also show how the standard Callan-Symanzik equation applies also to the running of the initial conditions.\footnote{An unrelated but quite interesting appearance of a Callan-Symanzik equation in an inflationary setting occurs for a flow within the space of inflationary models \cite{mcnees}.}  Section~\ref{conclude} concludes with a brief outline of how initial state effects must be treated to address correctly the question of back-reaction as well as to calculate the expected trans-Planckian signal in the primordial power spectrum.

\section{Boundary conditions in an expanding space-time}
\label{frwbound}

We begin with the action for a free scalar field propagating in classical curved background,
\begin{equation}
S = \int d^4x\sqrt{-g} \left[ 
{\textstyle{1\over 2}} g^{\mu\nu} \nabla_\mu\varphi \nabla_\nu\varphi 
- {\textstyle{1\over 2}} \xi R \varphi^2 
- {\textstyle{1\over 2}} m^2 \varphi^2 \right]
\label{action}
\end{equation}
where $g=\det(g_{\mu\nu})$ and $R$ is the scalar curvature.\footnote{Our convention for the signature of the Riemann curvature tensor is defined by $R^\lambda_{\ \mu\nu\rho} = \partial_\rho \Gamma^\lambda_{\mu\nu} - \partial_\nu \Gamma^\lambda_{\mu\rho} + \Gamma^\lambda_{\rho\sigma} \Gamma^\sigma_{\mu\nu} - \Gamma^\lambda_{\nu\sigma} \Gamma^\sigma_{\mu\rho}$ and the scalar curvature corresponds to $R = g^{\mu\nu} R^\lambda_{\ \mu\lambda\nu}$.}  On very large scales or at early times, the universe appears highly homogeneous and isotropic so we shall consider backgrounds where the metric only depends on time, 
\begin{equation}
ds^2 = dt^2 - a^2(t)\, d\vec x\cdot d\vec x , 
\label{frwmetrict}
\end{equation}
although the setting can be readily generalized to less symmetric backgrounds as well.  This general Robertson-Walker metric can also be expressed in a conformally flat form, 
\begin{equation}
ds^2 = g_{\mu\nu}\, dx^\mu dx^\nu 
= a^2(\eta) \left[ d\eta^2 - d\vec x\cdot d\vec x \right] , 
\label{frwmetric}
\end{equation}
by defining a conformal time with 
\begin{equation}
\eta(t) = \int^t_0 {dt'\over a(t')}  
\label{confdef}
\end{equation}
and setting $a(\eta)=a(\eta(t))$.  We shall work in this conformal coordinate system since the conformal time has a useful physical interpretation.  In units where $c=1$, the conformal time corresponds to the distance traveled by a massless particle since the earliest of times.  Thus, simultaneous points separated by a spatial distance greater than $\eta$ were never in causal contact.  The conformal time is moreover used in the standard inflationary calculations of the primordial power spectrum.

The simplest curved background of this Robertson-Walker form is de Sitter space, 
\begin{equation}
a(\eta) \to {1\over H\eta} , 
\qquad
ds^2 = {d\eta^2 - d\vec x\cdot d\vec x\over H^2\eta^2} . 
\label{inflflat}
\end{equation}
The curvature of de Sitter space is everywhere constant, $R=12H^2$, so that in this case the mass term and the curvature term in the action are redundant and the curvature term can be absorbed by a suitable rescaling of the mass.  de Sitter space is sometimes used as an idealization of the conditions expected to occur during inflation, where the background curvature, while not constant, varies only slowly over time.  

We shall consider here a completely general isotropically expanding background.  In such a background, the rate of expansion is characterized by the Hubble parameter, 
\begin{equation}
H(\eta) = {a'\over a} \equiv {1\over a} {\partial a\over\partial\eta}, 
\label{Htdef}
\end{equation}
in terms of which the scalar curvature is 
\begin{equation}
R(\eta) \equiv {6\over a^2} \bigl[ H' + H^2 \bigr] 
= {6\over a^2} {a^{\prime\prime}\over a} . 
\label{Rval}
\end{equation}

Varying the action with respect to the field yields the Klein-Gordon equation, 
\begin{equation}
\nabla^2 \varphi + \xi R \varphi + m^2 \varphi =  0 . 
\label{KGfull}
\end{equation}
Since the spatial part of the metric is flat, the spatial eigenmodes are plane waves so that the expansion of the field, as a linear sum of creation and annihilation operators, can be expressed as  
\begin{equation}
\varphi(\eta,\vec x) = \int {d^3\vec k\over (2\pi)^3}\, \left[ 
U_k(\eta) e^{i\vec k\cdot\vec x} a_{\vec k}
+ U_k^*(\eta) e^{-i\vec k\cdot\vec x} a^\dagger_{\vec k}
\right] . 
\label{opexpand}
\end{equation}
In Minkowski space, the time-dependent eigenmodes, $U_k(\eta)$, are also simple exponential functions but in an expanding background these mode functions are instead determined by 
\begin{equation}
U_k^{\prime\prime} + 2 H U_k^\prime 
+ \left[ k^2 + \xi a^2 R + a^2 m^2 \right] U_k = 0 
\label{KGRM}
\end{equation}
where $k\equiv |\vec k|$.  The solutions to this Klein-Gordon equation are completely determined once we have specified two constants of integration; actually, one of these is already fixed by the equal time commutation relation.  If the creation and annihilation operators are normalized to satisfy 
\begin{equation}
\bigl[ a_{\vec k}, a_{\vec k'}^\dagger \bigr] 
= (2\pi)^3 \delta^3(\vec k-\vec k') , 
\label{aadag}
\end{equation}
and the standard equal time commutation relation holds between the field $\varphi(x)$ and its conjugate momentum $\pi(x)$, 
\begin{equation}
\bigl[ \pi(\eta,\vec x), \varphi(\eta,\vec y) \bigr] 
= - i \delta^3(\vec x-\vec y) , 
\qquad
\pi = a^2\, \varphi' , 
\label{etcrel}
\end{equation}
then the mode functions must satisfy the following Wronskian condition, 
\begin{equation}
a^2 \bigl[ U_k \partial_\eta U_k^* - U_k^* \partial_\eta U_k \bigr] = i .
\label{wronk}
\end{equation}
The second constant of integration is then fixed by some assumption about the state that is appropriate for the physical setting being examined.

\subsection{Initial states}

In the usual calculation of the primordial spectrum of perturbations, the state chosen during inflation is assumed to be the vacuum.  In Minkowski space this statement is completely unambiguous and is little affected by the evolution of the state over time.  For example, we could fix the state at some initial time to be the lowest energy eigenstate for the free Hamiltonian.  This condition can be expressed more explicitly either by simply stating that the modes $U_k^{\rm flat}(t)$ are the positive energy eigenmodes, 
\begin{equation}
U_k^{\rm flat}(t) \propto {e^{-ikt}\over\sqrt{2k}} , 
\label{flateigenA}
\end{equation}
up to an arbitrary phase, or equivalently by establishing an initial condition on the modes of the form
\begin{equation}
{\partial\over\partial t} U_k^{\rm flat}(t) \Bigr|_{t=t_0} 
= - ik U_k^{\rm flat}(t_0) . 
\label{flateigenB}
\end{equation}
Here we have neglected the effects of the mass since in inflation it is generally assumed to be small.  If the free theory begins to break down at some scale $M$, so that the free Klein-Gordon equation receives nonnegligible corrections at this scale, the modes used in Eq.~(\ref{flateigenA}) might not be the correct eigenmodes of the full Klein-Gordon equation for $k>M$.  In terms of the differential form of the initial condition in Eq.~(\ref{flateigenB}), the simple constant of proportionality $-ik$ could receive more general corrections such as those which scale as $k/M$ to some power.  As long as we only evaluate processes at scales much lower than $M$, the details of the state above $M$ should be unimportant, since in Minkowski space there is no evolution of scales---what we mean by a small momentum compared to $M$, once specified, remains fixed at all times.

The absence of these two principles which held in flat space---the existence of a conserved Hamiltonian and the time-independence of scales---affects how we choose an appropriate state for inflation.  At very short intervals and over brief changes in time, the curvature of the background is not apparent and so we can choose our modes so that they match with the Minkowski modes in this regime.  More specifically, the leading behavior of the solution to the Klein-Gordon equation (\ref{KGRM}) at short distances, $k\gg a\sqrt{R}, am$, is described by 
\begin{equation}
U_k(\eta) = c_k\, {e^{-ik(\eta-\eta_0)}\over a(\eta)\sqrt{2k}} 
+ d_k\, {e^{ik(\eta-\eta_0)}\over a(\eta)\sqrt{2k}} 
+ \cdots . 
\label{BDlimit}
\end{equation} 
If we define our state again at some initial time, $\eta=\eta_0$, rescaling coordinates so that 
\begin{equation}
a(\eta_0) = 1 , 
\label{aeta0def}
\end{equation}
then over short intervals---such that $(\eta-\eta_0)H(\eta_0) \ll 1$---away from this spacelike surface we can also neglect the redshifting of the scales.  In this limit, the change in conformal time is essentially the same as the change in cosmic time, 
\begin{equation}
\eta-\eta_0 = {t-t_0\over a(t_0)} \left[ 
1 - {1\over 2} {\dot a(t_0)\over a(t_0)} (t-t_0) + \cdots \right] 
\approx t-t_0 ,
\label{scalenear0}
\end{equation}
so that the mode functions become the positive and negative energy modes of Minkowski space, 
\begin{equation}
U_k(\eta\approx\eta_0) = c_k {e^{-ik(\eta-\eta_0)}\over\sqrt{2k}} 
+ d_k {e^{ik(\eta-\eta_0)}\over\sqrt{2k}} 
+ \cdots . 
\label{BDlimitnear}
\end{equation} 
We then obtain the usual vacuum structure at intervals where the background curvature is not noticeable by setting $d_k=0$.

This prescription defines the Bunch-Davies \cite{bunch} vacuum state, $|E\rangle$.\footnote{Our notation is adopted from that of de Sitter space where this state is also frequently called the Euclidean vacuum, since it is the unique invariant state that is regular when analytically continued to lower half of the Euclidean sphere.}  We shall write the modes associated with this state as $U_k^E(\eta)$.  Just as in the case of the Minkowski vacuum above, we can specify the state either by simply stating that its modes are asymptotically proportional to the Minkowski vacuum modes, 
\begin{equation}
U_k^E(\eta) \to {e^{-ik(\eta-\eta_0)}\over a(\eta)\sqrt{2k}} 
\qquad{\rm as}\ k\to\infty , 
\label{frweigenA}
\end{equation}
or by a differential initial condition on the modes, 
\begin{equation}
\nabla_n U_k^E(\eta_0) = - i \omega_k[\eta_0] U_k^E(\eta_0) . 
\label{frweigenB}
\end{equation}
The differential operator here is the covariant derivative in the direction normal to the initial surface,
\begin{equation}
\nabla_n \equiv n^\mu\nabla_\mu . 
\label{nabndef}
\end{equation}
For a simple spacelike surface such as $\eta=\eta_0$, the unit normal to the surface is $n_\mu = \bigl( a(\eta), \vec 0 \big)$.  We shall frequently abbreviate $\omega_k[\eta_0]$ by $\omega_k$, but it is important to remember that it can depend on the initial time, 
\begin{equation}
\omega_k[\eta_0] 
= {i\partial_\eta U^E_k(\eta_0)\over a(\eta_0)U^E_k(\eta_0)} . 
\label{BDinomega}
\end{equation}
When considering more general initial states it will be useful to choose a particular phase convention for the Bunch-Davies modes, so we hereafter shall let
\begin{equation}
U_k^E(\eta_0) = U_k^{E*}(\eta_0) . 
\label{phase}
\end{equation}

Although we have chosen the state by considering the asymptotic behavior of the modes at large momenta, $k\gg a\sqrt{R}$, as we continue to yet shorter distances we encounter the same possibility mentioned before for flat space---that the free Klein-Gordon equation could receive substantial corrections for $k\sim M$ and above.  For example, $M$ could correspond to the scale at which some new dynamics becomes strongly interacting with the inflaton or it could represent the scale at which the classical description of gravity breaks down.

To include such effects, let us consider a more general initial state determined by the boundary condition, 
\begin{equation}
\nabla_n U_k(\eta_0) = -i\varpi_k[\eta_0]\, U_k(\eta_0) . 
\label{incond}
\end{equation}
From a low energy perspective, where we assume that the Bunch-Davies modes describe the ``vacuum'' extrapolated to arbitrarily short distances, it is convenient to express the modes for this general state as a transformation of the Bunch-Davies modes, 
\begin{equation}
U_k(\eta) = N_k \left[ U_k^E(\eta) + e^{\alpha_k} U_k^{E*}(\eta) \right] , 
\label{genmode}
\end{equation}
where the {\it initial state structure function\/} $e^{\alpha_k}$ is 
\begin{equation}
e^{\alpha_k} = {\omega_k-\varpi_k\over \omega_k^*+\varpi_k} . 
\label{alphadef}
\end{equation}
Both the general modes and the Bunch-Davies modes obey the Wronskian condition given earlier in Eq.~(\ref{wronk}), so the normalization is completely determined by $e^{\alpha_k}$, up to an arbitrary phase, 
\begin{equation}
N_k = {1\over\sqrt{1-e^{\alpha_k+\alpha_k^*}}} . 
\label{Nadef}
\end{equation}
Note that although we have chosen a particular phase for the Bunch-Davies modes with Eq.~(\ref{phase}), we can always choose an arbitrary relative phase between the two terms in a general mode by suitably choosing the phase of $e^{\alpha_k}$.

The structure function $e^{\alpha_k}$ describes how the state differs from the assumed vacuum at different scales.  At very large distances, if we are considering an excited state the structure function need not vanish; but the signals of new physics should not be very apparent since the approximation that the theory is that of a nearly free scalar field is good far below $M$.  In this regime, it is natural for the effects of new physics to be suppressed by powers of $k/M$. 

Our goal here is to implement an effective theory description of the initial state.  From this perspective, the state in Eq.~(\ref{genmode}) is only meant to be appropriate for observables measured at scales well below $M$, and not that for a complete theory which is applicable to measurements made at any scale.  As a consequence, the effective states can contain structures which are the analogues of the nonrenormalizable operators used in an effective field theory Lagrangian.  In both cases, the theory remains predictive at long distances since there is a natural small parameter given by the ratio of the energy or momentum of the process being studied to the scale of new physics $M$.  In both cases too, renormalization of the theory can introduce further higher order corrections so that an infinite number of constants is often needed to make a prediction to arbitrary accuracy; but to any finite accuracy, only a small number are needed since the rest are suppressed by high powers of the small ratio of scales.  At scales near $M$, the effective Lagrangian description breaks down but at these energies we should be able to observe the dynamics which produced the nonrenormalizable operators in the low energy effective theory.  Similarly, once we probe short distances directly, we should see corrections to the Klein-Gordon equation and the modes $U_k(\eta)$ given in Eq.~(\ref{genmode}) should be replaced with the correct short-distance eigenmodes.

The remaining significant departure from Minkowski space is the constant redshifting of scales inherent to an expanding background, which is responsible for the trans-Planckian problem.  The effective theory description of the initial state relies upon the smallness of the measured scale, $k_{\rm exp}$, compared with the scale of new physics, $M$, but this ratio is also influenced by the expansion, 
\begin{equation}
{a(\eta_{\rm now})\over a(\eta_0)} {k_{\rm exp}\over M} \ll 1 , 
\label{limiteta}
\end{equation}
and the earliest time for which perturbative calculation works is one which does quite saturate this bound,
\begin{equation}
{a(\eta_0^{\rm earliest})\over a(\eta_{\rm now})} 
\sim {k_{\rm exp}\over M} . 
\label{earlyeta}
\end{equation}
Although this time dependence of scales limits the applicability of the effective theory, it should not be seen as anything mysterious or that $\eta_0$ must be chosen either at this bound or at a time when some nontrivial dynamics is occurring.  To study the inflationary prediction for the cosmic microwave background power spectrum, for example, it is sufficient to choose an ``initial time'' when all of the features of the currently observed power spectrum are just within the horizon during inflation and which still satisfies the condition in Eq.~(\ref{limiteta}) for a well behaved perturbation theory.  What the effective theory approach accomplishes is not a complete description of the theory to an arbitrarily early time, but rather it provides a completely generic parameterization of the effects of these earlier epochs or of higher scale physics once the state has entered a regime where they can be treated perturbatively.

\subsection{Adiabatic modes}

Solving for the analytic form of the eigenmode functions in a completely general Robertson-Walker background is frequently not possible, so instead we must find a consistent method for approximating the modes.  A standard approach for approximating the Bunch-Davies modes is provided by the {\it adiabatic modes\/}.  The adiabaticity here refers to assuming that the time derivatives are small compared to the scales we are examining.  For our purpose of renormalizing the theory, we usually need to know the detailed form for the modes for large momenta, while the exact dependence at longer wavelengths, which is important for detailed finite effects, does not affect the divergences or the accompanying running of the parameters of the theory.

We begin by writing the Bunch-Davies mode functions in a form that superficially resembles that of the flat space modes, 
\begin{equation}
U_k^E(\eta) = {e^{-i\int_{\eta_0}^\eta d\eta'\, \Omega_k(\eta')}\over a(\eta) \sqrt{2\Omega_k(\eta)}} . 
\label{wkbmode}
\end{equation}
The {\it generalized frequency\/} function $\Omega_k(\eta)$ is determined by the differential equation, 
\begin{equation}
\Omega_k^2 = k^2 + a^2 m^2 
+ \Bigl[ \xi - {1\over 6} \Bigr] a^2 R 
- {1\over 2} {\Omega_k^{\prime\prime}\over\Omega_k}
+ {3\over 4} {\Omega_k^{\prime 2}\over\Omega_k^2} , 
\label{bigOmegadef}
\end{equation}
which is derived by substituting the modes in Eq.~(\ref{wkbmode}) into the Klein-Gordon equation, Eq.~(\ref{KGRM}).  We shall see by the end of this section that these modes indeed satisfy the Bunch-Davies condition.

In the adiabatic approximation, time derivatives, whether of the scale factor $a(\eta)$ or of the generalized frequency $\Omega_k(\eta)$, are small.  This assumption allows Eq.~(\ref{bigOmegadef}) to be solved through a series of successively better approximations, 
\begin{equation}
\Omega_k^2(\eta) = [\Omega_k^{(0)}(\eta)]^2 
+ [\Omega_k^{(2)}(\eta)]^2 + \cdots ,
\label{bigOmegaIT}
\end{equation}
starting at zeroth order with 
\begin{equation}
\Omega_k^{(0)}(\eta) = \sqrt{ k^2 + a^2(\eta)m^2 } ,
\label{bigOmega0}
\end{equation}
and then proceeding iteratively, using a lower order solution to fix the next order,
\begin{equation}
[\Omega_k^{(2)}]^2
= \biggl[ \xi - {1\over 6} \biggr] R 
- {1\over 2} {\Omega_k^{(0)\prime\prime}\over\Omega_k^{(0)}}
+ {3\over 4} \biggl[ {\Omega_k^{(0)\prime}\over\Omega_k^{(0)}} \biggr]^2 . 
\label{bigOmega2}
\end{equation}
In particular, inserting the zeroth order expression into this equation yields
\begin{eqnarray}
[\Omega_k^{(2)}]^2
&=& \Bigl[ \xi - {1\over 6} \Bigr] R 
- {1\over 2} {(H' + 2H^2)a^2m^2\over k^2 + a^2m^2} 
\nonumber \\
&&
+ {5\over 4} {H^2 a^4m^4\over (k^2 + a^2m^2)^2} . \quad
\label{bigOmega2A}
\end{eqnarray}
The form of the generalized frequency becomes a little simpler if we retain only the leading terms in the limit, $k\gg am$; to second order in the adiabatic solution we have
\begin{equation}
\Omega_k^2(\eta) = \bigl[ k^2 + a^2(\eta){\cal M}^2(\eta) \bigr] 
+ \cdots , 
\label{bigOmega2AUV}
\end{equation}
where we have introduced an effective, time-dependent mass defined by 
\begin{equation}
{\cal M}^2(\eta) \equiv m^2
+ \Bigl[ \xi - {1\over 6} \Bigr] R(\eta) . 
\label{calMdef}
\end{equation}
Note that in the extreme ultraviolet limit, the leading behavior simplifies to 
\begin{equation}
\Omega_k(\eta) \approx k  
\label{bigOmegaXUV}
\end{equation}
reproducing the correct asymptotic behavior of a Bunch-Davies mode, as in Eq.~(\ref{frweigenA}).

\section{Propagation}
\label{prop}

In a pure Bunch-Davies state, the propagator straightforwardly generalizes the Feynman propagator for the Minkowski vacuum, 
\begin{eqnarray}
-iG_F^E(x,x') &=& \Theta(\eta-\eta')\, 
\langle E | \varphi(x) \varphi(x') | E \rangle 
\nonumber \\
&&
+\ \Theta(\eta'-\eta)\, 
\langle E | \varphi(x') \varphi(x) | E \rangle . 
\qquad
\label{BDprop}
\end{eqnarray}
From the perspective of the low energy theory, this initial state is empty of all information about any higher scale physics, so this Green's function only propagates the information associated with sources moving through the bulk of space-time and no separate information propagates from the initial surface,
\begin{equation}
\bigl[ \nabla_x^2 + m^2 \bigr] G_F^E(x,x') 
= {\delta^4(x-x')\over\sqrt{-g(x)}} . 
\label{BDsource}
\end{equation}
For a more general initial state, the Green's function will need also to include consistently the propagation of this initial state information.

One way to express this consistency is to impose the same boundary condition on the propagator as that which determined the mode functions.  For the Bunch-Davies propagator, written in its spatial momentum representation, 
 \begin{eqnarray}
G_F^E(x,x') = \int {d^3\vec k\over (2\pi)^3}\, 
e^{i\vec k\cdot(\vec x-\vec x')} G_k^E(\eta,\eta')
\label{BDpropFT}
\end{eqnarray}
with 
\begin{eqnarray}
-i G_k^E(\eta,\eta') 
&=& \Theta(\eta-\eta')\, U_k^E(\eta) U_k^{E*}(\eta')
\nonumber \\
&&
+\ \Theta(\eta'-\eta)\, U_k^{E*}(\eta) U_k^E(\eta') , 
\label{BDkprop}
\end{eqnarray}
we find that in the physical region---those times subsequent to the initial time---this propagator is consistent with a boundary condition for each of its arguments, 
\begin{eqnarray}
n^\mu\nabla_\mu G_k^E(\eta,\eta')\bigr|_{{\eta=\eta_0\atop \eta'>\eta_0}} 
&=& i\omega_k^* G_k^E(\eta_0,\eta')\bigr|_{\eta'>\eta_0} 
\nonumber \\
n^\mu\nabla'_\mu G_k^E(\eta,\eta')\bigr|_{{\eta'=\eta_0\atop \eta>\eta_0}} 
&=& i\omega_k^* G_k^E(\eta,\eta_0)\bigr|_{\eta>\eta_0} . 
\label{GkBDbnd}
\end{eqnarray}
Note that the right side is the complex conjugate of the coefficient of the boundary condition defined for the modes.  The reason is that for $\eta=\eta_0$ and $\eta'>\eta_0$, for example, the non-vanishing  $\Theta$-function is that accompanying the $U_k^{E*}(\eta)U_k^E(\eta')$ factor and not its conjugate, so the time-derivative acts upon $U_k^{E*}(\eta)$.

The nontrivial new element is that the normal derivatives also act on the $\Theta$-functions associated with the time-ordering.  In the Bunch-Davies propagator, the opposite ordering of the arguments in the two $\Theta$-functions causes the resulting $\delta$-function terms to cancel between its forward and backward propagating terms.  In fact, the boundary condition in Eq.~(\ref{GkBDbnd}) is not quite sufficient to determine completely the propagator, but choosing the propagator to be locally time-translationally invariant over infinitesimally short intervals fixes the remaining ambiguity.

To understand how the propagator is determined in more detail, we shall make a short excursion into Minkowski space.  There we can write a propagator very generally as 
\begin{equation}
\tilde G_k(t,t') 
= \Theta(t-t')\, \tilde G_k^>(t,t') + \Theta(t'-t)\, \tilde G_k^<(t,t') 
\label{tidleGdef}
\end{equation}
where we shall use tildes to denote the propagator in flat space described by the coordinates $(t,\vec x)$.  The Wightman functions $\tilde G_k^{>,<}(t,t')$ are partially fixed by three conditions:  the propagator should be continuous at $t=t'$,
\begin{equation}
\tilde G_k^>(t,t) = \tilde G_k^<(t,t) ,
\label{cont}
\end{equation}
it should satisfy the correct discontinuity in its first derivative to produce a correctly weighted point-source, 
\begin{equation}
\bigl[ \partial_t \tilde G_k^>(t,t') 
- \partial_t \tilde G_k^<(t,t') \bigr]_{t'=t} = 1 ,
\label{jump}
\end{equation}
and away from the point-source, it should satisfy the Klein-Gordon equation, 
\begin{equation}
\Bigl[ {d^2\over dt^2} + k^2 \Bigr] \tilde G_k^{>,<}(t,t') = 0 
= \Bigl[ {d^2\over dt^{\prime 2}} + k^2 \Bigr] \tilde G_k^{>,<}(t,t') . 
\label{MKG}
\end{equation}
To avoid any possible confusion with the $\omega_k$ defined earlier, we have set the mass to zero for this example.  Applying all three conditions gives 
\begin{eqnarray}
\tilde G_k^>(t,t') 
&=& \Bigl[ {i\over 2k} + a_k \Bigr] e^{-ik(t-t')} + b_k e^{ik(t-t')} 
\nonumber \\
&&
+ c_k e^{-ik(t+t')} + d_k e^{ik(t+t')} 
\nonumber \\
\tilde G_k^<(t,t') 
&=& a_k e^{-ik(t-t')} + \Bigl[ {i\over 2k} + b_k \Bigr] e^{ik(t-t')} 
\nonumber \\
&&
+ c_k e^{-ik(t+t')} + d_k e^{ik(t+t')} . 
\label{3confl}
\end{eqnarray}
If we further apply a boundary condition at $t_0$ analogous to the Bunch-Davies condition above, 
\begin{eqnarray}
\partial_t \tilde G_k^E(t,t')\bigr|_{t=t_0,\ t'>t_0} 
&=& ik \tilde G_k^E(t_0,t')\bigr|_{t'>t_0} 
\nonumber \\
\partial_{t'} \tilde G_k^E(t,t')\bigr|_{t'=t_0,\ t>t_0} 
&=& ik \tilde G_k^E(t,t_0)\bigr|_{t>t_0} , 
\label{GkBDbndM}
\end{eqnarray}
we very nearly obtain the correct Feynman propagator, 
\begin{eqnarray}
\tilde G_k^{E,>}(t,t') 
&=& {i\over 2k} e^{-ik(t-t')} + d_k e^{ik(t+t')} 
\nonumber \\
\tilde G_k^{E,<}(t,t') 
&=& {i\over 2k} e^{ik(t-t')} + d_k e^{ik(t+t')} . 
\label{4confl}
\end{eqnarray}
The remaining condition we impose is that the propagator should be time-translationally invariant so that it should only depend on $t-t'$ and not $t+t'$ since the boundary condition in Eq.~(\ref{GkBDbndM}) does not itself break time translation invariance.  This last condition requires $d_k=0$ so that the standard Minkowski space propagator is obtained.

Now consider a boundary condition in Minkowski space which explicitly breaks the time-translation invariance, 
\begin{eqnarray}
\partial_t \tilde G_k(t,t')\bigr|_{t=t_0,\ t'>t_0} 
&=& i\kappa \tilde G_k(t_0,t')\bigr|_{t'>t_0} 
\nonumber \\
\partial_{t'} \tilde G_k(t,t')\bigr|_{t'=t_0,\ t>t_0} 
&=& i\kappa \tilde G_k(t,t_0)\bigr|_{t>t_0} ; 
\label{GkAbndM}
\end{eqnarray}
applying this condition yields instead 
\begin{eqnarray}
\tilde G_k^>(t,t') 
&=& 
\Bigl[ {i\over 2k} + b_k \Bigr] 
\bigl[ e^{-ik(t-t')} + e^{\tilde\alpha_k} e^{ik(2t_0-t-t')} \bigr] 
\nonumber \\
&&
+ b_k \bigl[ e^{ik(t-t')} + e^{-\tilde\alpha_k} e^{-ik(2t_0-t-t')} \bigr] 
\nonumber \\
\tilde G_k^<(t,t') 
&=& 
\Bigl[ {i\over 2k} + b_k \Bigr] \bigl[ e^{ik(t-t')} + e^{\tilde\alpha_k} e^{ik(2t_0-t-t')} \bigr] 
\nonumber \\
&&
+ b_k \bigl[ e^{-ik(t-t')} + e^{-\tilde\alpha_k} e^{-ik(2t_0-t-t')} \bigr] 
\label{4conflA}
\end{eqnarray}
where we have defined the Minkowski space initial state structure function by \begin{equation}
e^{\tilde\alpha_k} \equiv {k-\kappa\over k+\kappa} . 
\label{Missf}
\end{equation}
Sending $\kappa\to k$ sends $e^{\tilde\alpha_k}\to 0$ so we should recover the vacuum propagator in this limit.  This requirement fixes the remaining ambiguity in a propagator consistent with a general initial state by setting $b_k=0$,\footnote{Actually, we could have a coefficient $b_k$ that only diminishes faster than $e^{\tilde\alpha_k}$, but we exclude this possibility by demanding that a perturbation theory based on our propagator should be free of uncontrolled divergences, such as those \cite{fate,einhorn,banks,lowe,taming} occurring in the $\alpha$-vacua of de Sitter space \cite{alpha}, which would occur for nonzero values of $b_k$.} 
\begin{eqnarray}
\tilde G_k(t,t') 
&=& \Theta(t-t')\, {i\over 2k} e^{-ik(t-t')} 
+ \Theta(t'-t)\, {i\over 2k} e^{ik(t-t')} 
\nonumber \\
&&
+ {i\over 2k} e^{\tilde\alpha_k} e^{ik(2t_0-t-t')} . 
\label{5conflA}
\end{eqnarray}
Sometimes it is convenient to write this propagator in a more suggestive form by defining an image time, $t_{{\scriptscriptstyle I}} \equiv 2t_0 - t$, 
\begin{eqnarray}
\tilde G_k(t,t') 
&=& \Theta(t-t')\, {i\over 2k} e^{-ik(t-t')} 
+ \Theta(t'-t)\, {i\over 2k} e^{ik(t-t')} 
\nonumber \\
&&
+ \Theta(t_{{\scriptscriptstyle I}}-t')\, {i\over 2k} e^{\tilde\alpha_k} e^{-ik(t_{{\scriptscriptstyle I}} -t')} 
\nonumber \\
&&
+ \Theta(t'-t_{{\scriptscriptstyle I}})\, {i\over 2k} e^{\tilde\alpha_k} e^{ik(t_{{\scriptscriptstyle I}} -t')} . 
\label{5conflAim}
\end{eqnarray}
Since the theory is only applicable to the region subsequent to the initial surface, $t,t'>t_0$ this propagator always agrees with Eq.~(\ref{5conflA}) in this region and is moreover still consistent with the boundary condition in Eq.~(\ref{GkAbndM}).  In this form, the initial state propagator contains two sources, one for the physics point-source and one for a fictitious image source which encodes the initial state information.

Returning to a general expanding background, if we impose the analogous conditions on the propagator---continuity, an appropriate jump in its first derivative for a point source, consistency with the Klein-Gordon equation---as well as a general initial condition, 
\begin{eqnarray}
n^\mu\nabla_\mu G_k^\alpha(\eta,\eta')\bigr|_{{\eta=\eta_0\atop \eta'>\eta_0}} 
&=& i\varpi_k^* G_k^\alpha(\eta_0,\eta')\bigr|_{\eta'>\eta_0} 
\nonumber \\
n^\mu\nabla'_\mu G_k^\alpha(\eta,\eta')\bigr|_{{\eta'=\eta_0\atop \eta>\eta_0}} 
&=& i\varpi_k^* G_k^\alpha(\eta,\eta_0)\bigr|_{\eta>\eta_0} , 
\label{Gkbnd}
\end{eqnarray}
then we obtain a unique propagator structure,
\begin{eqnarray}
-iG_k^\alpha(\eta,\eta') 
&=& \Theta(\eta-\eta')\, U_k^E(\eta) U_k^{E*}(\eta') 
\nonumber \\
&&
+\ \Theta(\eta'-\eta)\, U_k^{E*}(\eta) U_k^E(\eta')
\nonumber \\
&&
+\ e^{\alpha_k^*} U_k^E(\eta) U_k^E(\eta') , 
\label{GAprop}
\end{eqnarray}
which matches with the Bunch-Davies propagator in the limit where the initial state structure function vanishes.

The origin of this propagator has an elegant interpretation as the generalization of the time-ordering in the usual vacuum propagator.  Recall that in Minkowski space the time-ordering produces the forward propagation of positive frequencies and the backward propagation of negative frequencies.  If we start with a different initial state, we must also account for propagation of the initial state information as well.  This propagation from the boundary should contain only the forward propagation of one set of modes since the backward propagation would be into the region before $\eta_0$.  Thus we have only the $U_k^E(\eta)U_k^E(\eta')$ term in the propagator and not its complex conjugate, which would represent the backwards propagating negative frequency modes associated with the initial state.  This interpretation becomes clearer when we rewrite the propagator as 
\begin{eqnarray}
-iG_k^\alpha(\eta,\eta') 
&\propto& \Theta(\eta-\eta') U_k^E(\eta) U_k^*(\eta') 
\nonumber \\
&&
+\ \Theta(\eta'-\eta) U_k^*(\eta) U_k^E(\eta') . 
\label{Galphalt}
\end{eqnarray}
Its asymmetric form reflects the fact that signals from the point-source propagate both forward and backwards in time---depending upon the energy of the modes---while the boundary effects only propagate forward, since we never evaluate times earlier than the time at which the conditions are imposed.

\section{The renormalization condition}
\label{renormalization}

In an interacting field theory it is extremely rare that the evolution of a system can be solved exactly, even in flat space, and what is done instead is to describe processes perturbatively.  If successive perturbative corrections are sufficiently convergent, then the theory can be predictive as long as the experimental error is larger than the error we make in truncating the series.  The energies and the momenta in all parts of the leading term in this series are usually completely finite, being fixed by the actual momenta of the external, physical fields being measured.  However, in all the higher corrections appear intermediate processes where a field ranges over all possible momenta, including arbitrarily large ones.  Summing over this arbitrarily large momentum behavior, or equivalently the short-distance features of the theory, can produce divergences.  But since these divergences are constant, not depending on any measurable quantity, they can be absorbed by a suitable rescaling of the parameters of the theory and in terms of these renormalized quantities, the terms of the perturbation series are finite at each order.  In the process, we lose the idea of constant parameters---masses or couplings---and the renormalized parameters depend now on the scale at which they are defined.  This scale dependence is not arbitrary, but is is fixed by {\it renormalization conditions\/} which express how or at what scale a particular physical parameter is defined.

When a field starts in an initial state which contains some structure at short distances that differs from the na\"\i ve idea of a vacuum based on extrapolating the free theory to arbitrary scales, the intermediate processes in the perturbation series also sum over all of this short-distance structure of the initial state which can produce divergences in addition to those associated with the properties of a field propagating through the bulk of space-time.  Since they are associated with the structure of the initial state, this class of divergences is localized at the initial time \cite{greens} and they are removed by renormalizing the theory at this initial boundary.  Once renormalized there, the theory remains finite for all subsequent times.  

Establishing that these state-dependent divergences can be renormalized is important even if we are only interested initially in evaluating the leading term in the perturbative expansion, where all intermediate momenta are finite.  Partially this importance lies in the fact that this leading result only has any meaning if the corrections to it are finite and are small compared to it.  But further, even the detailed form of the leading result depends on how the theory propagates forward the information contained in the initial state and whether this propagation is consistent is what is being checked when we renormalize the perturbative corrections.  Thus knowing how to renormalize the higher order corrections---for example in the two-point function---tells us also the correct form for the leading, tree-level prediction for the trans-Planckian signature in the cosmic microwave background.

For the $S$-matrix in Minkowski space, a standard set of renormalization conditions is applied, such as the vanishing of the one-particle expectation value or the location and the residue of the pole associated with the physical mass of the particle.  During inflation, the physical setting differs quite dramatically from that assumed for the $S$-matrix so the form of the renormalization conditions must be modified appropriately.  A typical inflationary model contains a scalar field, the inflaton, which we divide into a spatially independent classical zero mode $\phi(\eta)$ and a fluctuation $\psi(\eta,\vec x)$ about this value,
\begin{equation}
\varphi(\eta,\vec x) = \phi(\eta) + \psi(\eta,\vec x) . 
\label{fieldvev}
\end{equation}
The zero mode drives the overall inflationary era and its value changes slowly as the field rolls down its potential.  The fluctuations, combined with the scalar component of the fluctuations of the gravitational background, produce the pattern of nearly scale-invariant, nearly Gaussian primordial perturbations that seed the density perturbations seen both in the temperature fluctuations, observed by WMAP, and the observed large-scale structure, before nonlinear dynamics have set in.

Because the zero mode corresponds to the classical expectation value of the field, the vanishing of the expectation value of the fluctuations provides a natural renormalization condition for this setting \cite{neft,weinberg}, 
\begin{equation}
\langle \alpha_k(\eta) | \psi(x) | \alpha_k(\eta) \rangle = 0 . 
\label{renormcond}
\end{equation}
As we shall see below, because the fluctuation couples to the classical zero mode, this tadpole condition will allow us to define the renormalized mass and the coupling by refering to their values for the zero mode.

The method for constructing an effective description of a general initial condition described earlier applies to any interacting field theory so we shall illustrate the boundary renormalization with the relatively simple example provided by a quartic interaction, 
\begin{equation}
S = \int d^4x\, \sqrt{-g} \left[ {\textstyle{1\over 2}} 
g^{\mu\nu} \nabla_\mu\varphi \nabla_\nu\varphi 
- {\textstyle{1\over 2}} \xi R\varphi^2 
- {\textstyle{1\over 2}} m^2 \varphi^2 
- {\textstyle{1\over 24}} \lambda \varphi^4 \right] .  
\label{quarticaction}
\end{equation}
Decomposing the full field into its zero mode and fluctuating components yields
\begin{equation}
S = S_\phi + S_\psi^0 + S_\psi^{\rm int}
\label{actionphipsi}
\end{equation}
where $S_\phi$ is the classical action for the zero mode obtained by setting $\varphi\to\phi$ in Eq.~(\ref{quarticaction}) and $S_\psi^0$ and $S_\psi^{\rm int}$ are the free and the interacting parts of the action for the fluctuations, 
\begin{eqnarray}
S_\psi^0 &=& 
\int d^4x\, \sqrt{-g} \bigl\{ 
{\textstyle{1\over 2}} \nabla_\mu\psi \nabla^\mu\psi 
- {\textstyle{1\over 12}} R\psi^2 
\nonumber \\
&&\qquad\qquad\ 
- {\textstyle{1\over 2}} \left[
m^2 + {\textstyle{1\over 2}} \lambda \phi^2 
+ \left( \xi - {\textstyle{1\over 6}} \right) R
\right] \psi^2 
\bigr\} 
\qquad
\label{Spsifree}
\end{eqnarray}
and 
\begin{eqnarray}
S_\psi^{\rm int} &=& 
\int d^4x\, \sqrt{-g} \bigl[ 
- \left[ \nabla^2\phi + \xi R\phi + m^2 \phi  
+ {\textstyle{1\over 6}} \lambda \phi^3 \right] \psi 
\nonumber \\
&&\qquad\qquad\quad 
- {\textstyle{1\over 6}} \lambda \phi\psi^3 
- {\textstyle{1\over 24}} \lambda \psi^4 
\bigr] . 
\label{SpsiInt}
\end{eqnarray}

Notice that in expanding the interaction $\lambda\varphi^4$, we find a term that is quadratic in the fluctuation, $\lambda\phi^2\psi^2$, which acts as a time-dependent correction to the mass.  Therefore we have not included this term among the interactions but rather have grouped it with the other quadratic terms to form a time-dependent effective mass term $m^2 \to m^2 + {1\over 2}\lambda \phi^2(\eta)$.  We would have found the same shift had we treated this effect as an interaction once we summed the entire set of all possible insertions of this term in the free $\psi$ propagator.

The contribution of the zero mode to the effective mass similarly shifts the form of the adiabatic modes.  For example, the zeroth order adiabatic mode becomes,
\begin{equation}
[\Omega_k^{(0)}(\eta)]^2 = k^2 + a^2(\eta)\left( m^2 
+ {\textstyle{1\over 2}} \lambda\phi^2(\eta) \right) ,
\label{Om0shift}
\end{equation}
while the second order correction is now 
\begin{eqnarray}
[\Omega_k^{(2)}(\eta)]^2 
&=& \Bigl[ \xi - {1\over 6} \Bigr] a^2 R 
\nonumber \\
&&
- {1\over 2} {a^2 (H'+2H^2) \left( m^2 + {1\over 2}\lambda \phi^2 \right) 
\over k^2 + a^2 \left( m^2 + {1\over 2}\lambda \phi^2 \right) }
\nonumber \\
&&
- {\lambda\over 4} {\phi\phi^{\prime\prime} + \phi^{\prime 2}  + 4 H \phi\phi' 
\over k^2 + a^2 \left( m^2 + {1\over 2}\lambda \phi^2 \right) }
\nonumber \\
&&
+ {5\over 4} {a^4 \left[ H \left( m^2 + {1\over 2}\lambda \phi^2 \right) 
+ {1\over 2}\lambda \phi\phi' \right]^2
\over \left[ k^2 + a^2 \left( m^2 + {1\over 2}\lambda \phi^2 \right) \right]^2 } . \quad
\label{Om2shift}
\end{eqnarray}
However at very short distances, where the divergences in loop corrections occur, 
\begin{equation}
k^2 \gg a^2m^2, a^2\lambda\phi^2, 
{\lambda (\phi\phi^{\prime\prime} + \phi^{\prime 2} )\over R} , 
\label{kshortshift}
\end{equation}
the shift in the effective mass simply appears as a corresponding shift in the generalized mass defined earlier.  Thus the leading short-distance part of the generalized frequency to second order in the adiabatic expansion is still
\begin{equation}
\Omega_k(\eta) = \sqrt{ k^2 + a^2(\eta) {\cal M}^2(\eta) } + \cdots
\label{newOmlead}
\end{equation}
but with ${\cal M}^2(\eta)$ now given by 
\begin{equation}
{\cal M}^2(\eta) \equiv m^2 + {\textstyle{1\over 2}}\lambda\phi^2 
+ \left( \xi - {\textstyle{1\over 6}} \right) R . 
\label{newcalMdef}
\end{equation}

In the interaction picture, the free Hamiltonian determines the evolution of operators while the interaction Hamiltonian, $H_I$, determines the corresponding evolution of the states.  When the operator consists of a product of fields, the evolution is already contained in their time-dependence so the evolution of the tadpole from an initial state defined at $\eta=\eta_0$ to a later time $\eta_f$ is given in the Schwinger-Keldysh approach\footnote{A more detailed explanation of the Schwinger-Keldysh approach applied to this setting is given in the Appendix A of \cite{greens} which defines the notation used here and explains the correct time-ordering for contractions of any combination of $\psi^\pm(x)$ fields.} by 
\begin{eqnarray}
&&\!\!\!\!\!\!\!\!\!\!\!\!\!\!\!\!\!\!
\langle\alpha_k(\eta_f) | \psi^+(\eta_f,\vec x) | \alpha_k(\eta_f) \rangle 
\nonumber \\
&=& 
{\langle\alpha_k | T_\alpha \bigl( \psi^+(\eta_f,\vec x) 
e^{-i\int_{\eta_0}^0 d\eta\, \left[ H_I[\psi^+] - H_I[\psi^-] \right] }
\bigr) | \alpha_k \rangle \over 
\langle\alpha_k | T_\alpha \bigl( 
e^{-i\int_{\eta_0}^0 d\eta\, \left[ H_I[\psi^+] - H_I[\psi^-] \right] }
\bigr) | \alpha_k \rangle } \qquad
\label{psimatrix}
\end{eqnarray}
where $|\alpha_k\rangle = |\alpha_k(\eta_0)\rangle$.  Since both the $|\alpha_k \rangle$ state and $\langle\alpha_k|$ state have been time-evolved, we have two parts of the time evolution operator corresponding to each of these components.  The ``$+$'' fields are associated with the former while the ``$-$'' fields are associated with the latter and the relative minus sign between the two appearances of the interaction Hamiltonian is from the Hermitian conjugation of the unitary operator evolving the $\langle\alpha_k|$ state.

From the action in Eq.~(\ref{SpsiInt}), the interaction Hamiltonian for a simple quartic theory is 
\begin{eqnarray}
H_I[\psi^\pm] &=& 
\int d^3\vec y\, a^4 \bigl[ 
\left[ \nabla^2\phi + \xi R\phi + m^2 \phi  
+ {\textstyle{1\over 6}} \lambda \phi^3 \right] \psi^\pm 
\nonumber \\
&&\qquad\qquad
+ {\textstyle{1\over 6}} \lambda \phi\psi^{\pm 3} 
+ {\textstyle{1\over 24}} \lambda \psi^{\pm 4} 
\bigr] . 
\label{actHint}
\end{eqnarray}
which yields the following leading contribution to the tadpole matrix element, 
\begin{eqnarray}
0 &=& \langle \alpha_k(\eta_f) | \psi^+(x) | \alpha_k(\eta_f) \rangle
\nonumber \\
&=& - \int_{\eta_0}^{\eta_f} d\eta\, a^4(\eta) \int d^3\vec y\, 
\left[ G^>_0(x,y) - G^<_0(x,y) \right] 
\nonumber \\
&&\times
\Bigl\{ 
\nabla^2\phi(\eta) + \xi R(\eta)\phi(\eta) 
+ m^2\phi(\eta) + {\lambda\over 6} \phi^3(\eta) 
\nonumber \\
&&\qquad
- {i\lambda\over 2} \phi(\eta) G^>_\alpha(y,y) 
+ \cdots \Bigr\} 
\label{secondorder}
\end{eqnarray}
where $x=(\eta_f,\vec x)$, $y=(\eta,\vec y)$ and $z=(\eta',\vec z)$.  The Wightman functions, $G^{>,<}_\alpha(x,y)$, are defined by 
\begin{eqnarray}
G^>_\alpha(y,z) 
&=&
i\int {d^3\vec k\over (2\pi)^3}\, 
e^{i\vec k\cdot (\vec y-\vec z)}
\nonumber \\
&&\times
\left[ U_k^E(\eta) U_k^{E*}(\eta')
+ e^{\alpha_k^*} U_k^E(\eta) U_k^E(\eta')
\right] 
\nonumber \\
G^<_\alpha(y,z) 
&=& i\int {d^3\vec k\over (2\pi)^3}\, 
e^{i\vec k\cdot (\vec y-\vec z)}
\nonumber \\
&&\times
\left[ U_k^{E*}(\eta) U_k^E(\eta') 
+ e^{\alpha_k^*} U_k^E(\eta) U_k^E(\eta') 
\right] . \qquad
\label{aWights}
\end{eqnarray}
The Wightman functions labeled with a $0$ subscript correspond to those obtained by setting $e^{\alpha_k^*}=0$.  They appear here since in taking the difference $G^>_\alpha(x,y) - G^<_\alpha(x,y)$, the boundary-dependent terms of each function are the same and cancel each other.

\begin{figure}[!tbp]
\includegraphics{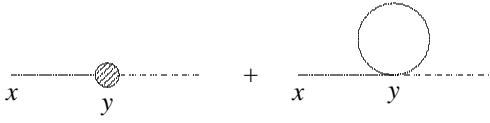}
\caption{The leading contributions to the running of the mass $m$ and the coupling $\lambda$ in a $\varphi^4$ theory.  The solid lines represent propagating $\psi$ fields while the dashed lines correspond to the zero mode $\phi$. 
\label{leadloops}}
\end{figure}

\begin{figure}[!tbp]
\includegraphics{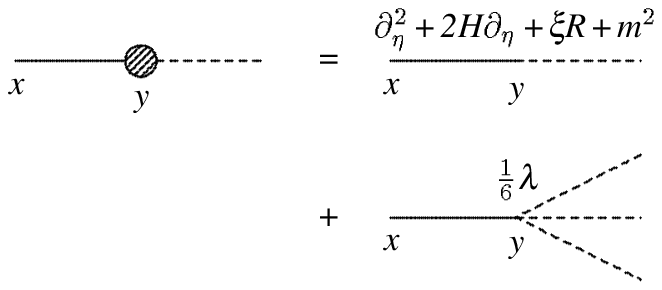}
\caption{The shaded blob corresponds to the following two graphs; note that the time derivatives act on the classical $\phi(\eta)$ field.
\label{opphi}}
\end{figure}

\begin{figure}[!tbp]
\includegraphics{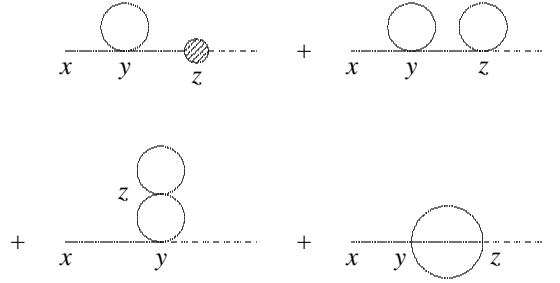}
\caption{Further graphs that contribute to the tadpole at second order.  The last of these graphs contains the leading nontrivial correction to the wavefunction renormalization.
\label{moreloops}}
\end{figure}

Diagrammatically, the leading contribution to the tadpole is shown in Figs.~\ref{leadloops}--\ref{opphi} where a $\psi$ propagator is represented by a solid line and a dashed line represents the zero mode $\phi$.  
At this order we shall encounter divergences which require the renormalization of the mass and the coupling of the field, but the leading contribution to the renormalization of the field only appears at two-loop-order through the last diagram shown in Fig.~\ref{moreloops}.

Substituting the expressions for the Green's functions in terms of the modes in Eq.~(\ref{aWights}) into the tadpole yields 
\begin{eqnarray}
&&\!\!\!\!\!\!\!\!\!\!\!\!\!\!\!\!\!\!\!\!\!\!\!\!
\langle \alpha_k(\eta_f) | \psi^+(x) | \alpha_k(\eta_f) \rangle
\nonumber \\
&=&
- \int_{\eta_0}^{\eta_f} d\eta\, a^4(\eta) 
{\cal G}(\eta_f,\eta) %
\nonumber \\
&&\times
\Bigl\{ 
\nabla^2\phi(\eta) + \xi R \phi(\eta) + m^2 \phi(\eta) 
+ {\lambda\over 6} \phi^3(\eta) 
\nonumber \\
&&\quad
+ {\lambda\over 2} \phi(\eta) \int {d^3\vec k\over (2\pi)^3}\, U_k^E(\eta) U_k^{E*}(\eta)
\nonumber \\
&&\quad
+ {\lambda\over 2} \phi(\eta) \int {d^3\vec k\over (2\pi)^3}\, 
e^{\alpha_k^*} U_k^E(\eta) U_k^E(\eta)
+ \cdots \Bigr\}
\label{leadingloops}
\end{eqnarray}
where the external $\psi$ leg has been abbreviated by 
\begin{equation}
{\cal G}(\eta_f,\eta) \equiv i \bigl[ 
U_0^E(\eta_f) U_0^{E*}(\eta) - U_0^{E*}(\eta_f) U_0^E(\eta) \bigr] . 
\label{calGdef}
\end{equation}
The external leg is independent of the initial state.  The first line within the braces in Eq.~(\ref{leadingloops}) is the equation of motion for the zero mode before we have included the corrections from its interactions with the fluctuations.

At leading order, the tadpole contains two terms with loop integrals, which appear in the last two lines of Eq.~(\ref{leadingloops}).  The first of these contains no dependence on the initial state; it produces the need to renormalize the mass and the coupling of the bulk $3+1$ dimensional field theory, as is shown in the next section.  In the second, the initial state structure function appears explicitly so all of the new divergences associated with the fine structure of the initial state arise at this order from this term.  Because of its importance in establishing the renormalizability of an effective description of a state bearing some trans-Planckian information, we shall treat it in much more detail separately, after first examining how the standard bulk renormalization proceeds in this inflationary setting.

\section{Bulk renormalization}
\label{bulk}

Through the coupling of the fluctuations to the zero mode, demanding that the one-point function for the fluctuations should vanish provides a simple and elegant origin for the renormalization and running of the bulk parameters:  $m$, $\lambda$, $\xi$.  Recall that the bare theory we have been considering, 
\begin{equation}
{\cal L} = 
{\textstyle{1\over 2}} g^{\mu\nu}\nabla_\mu\varphi \nabla_\nu\varphi
- {\textstyle{1\over 2}} \xi R \varphi^2 
- {\textstyle{1\over 2}} m^2 \varphi^2 
- {\textstyle{1\over 24}} \lambda \varphi^4 , 
\label{bareLag}
\end{equation}
can be also expressed in terms of the renormalized parameters by 
\begin{eqnarray}
{\cal L} 
&=& {\textstyle{1\over 2}} g^{\mu\nu}\nabla_\mu\varphi_R \nabla_\nu\varphi_R
- {\textstyle{1\over 2}} \xi_R R \varphi_R^2 
- {\textstyle{1\over 2}} m^2 \varphi_R^2 
- {\textstyle{1\over 24}} \lambda_R \varphi_R^4 
\nonumber \\
&& 
+\ {\textstyle{1\over 2}} (Z_3-1) g^{\mu\nu}\nabla_\mu\varphi_R \nabla_\nu\varphi_R
- {\textstyle{1\over 2}} (Z_\xi-1) \xi_R R \varphi_R^2 
\nonumber \\
&&
-\ {\textstyle{1\over 2}} (Z_0-1) m_R^2 \varphi_R^2 
- {\textstyle{1\over 24}} (Z_1-1) \lambda_R \varphi_R^4 . 
\label{renormLag}
\end{eqnarray}
The perturbative corrections to a Green's function calculated from the first line of this renormalized Lagrangian are very frequently divergent; these divergences arise from internal loops which sum over all possible momenta and contain a sufficiently small number of propagators.  Since these divergences appear only in the infinite momentum region of the loop integrals, they equivalently correspond to short-distance divergences and therefore can be cancelled by local counterterms of the form shown in the second two lines of Eq.~(\ref{renormLag}).  Choosing these counterterms correctly, the total value of the perturbative corrections at a given order is completely finite.

The bare and renormalized parameters are related to each other through a simple rescaling, 
\begin{eqnarray}
\varphi = Z_3^{1/2} \varphi_R
&\qquad&
\xi = {Z_\xi\over Z_3} \xi_R 
\nonumber \\
m^2 = {Z_0\over Z_3} m_R^2 
&\qquad&
\lambda = {Z_1\over Z_3^2} \lambda_R . 
\label{pararecs}
\end{eqnarray}
To leading order in the coupling $\lambda$, we only need to determine $Z_0$, $Z_1$ and $Z_\xi$ since the first correction to $Z_3$ only appears at two loop order, $\lambda^2$.  To control the divergences that appear in the perturbative corrections to a process, we must apply a regularization scheme such as dimensional regularization, which we use here.  

The bulk divergences occur at one loop order in the term 
\begin{equation}
{\lambda\over 2} \phi(\eta) \int {d^3\vec k\over (2\pi)^3}\, 
U_k^E(\eta) U_k^{E*}(\eta)
\label{bulkdivterm}
\end{equation}
appearing in the integrand of the tadpole in Eq.~(\ref{leadingloops}); here we have excised the external leg.  Substituting in the general form for the Bunch-Davies modes,
\begin{equation}
U_k^E(\eta) = {e^{-i\int_{\eta_0}^\eta d\eta'\, \Omega_k(\eta')}\over a(\eta) \sqrt{2\Omega_k(\eta)} } , 
\label{adiabatsenc}
\end{equation}
and applying the adiabatic approximation to second order to extract the leading behavior as $k\to\infty$ as in Eqs.~(\ref{newOmlead}--\ref{newcalMdef}), the divergent part of this loop is contained in 
\begin{equation}
{\lambda\over 4} {\phi(\eta)\over a^2(\eta)} \int {d^3\vec k\over (2\pi)^3}\, 
{1\over\sqrt{k^2+a^2(\eta){\cal M}^2(\eta)}} + \cdots . 
\label{bulkdivlead}
\end{equation}
By extending the number of spatial dimensions to $3-2\epsilon$, 
\begin{equation}
{\lambda\mu^{2\epsilon}\over 4} {\phi(\eta)\over a^2(\eta)} \int {d^{3-2\epsilon}\vec k\over (2\pi)^3}\, 
{1\over\sqrt{k^2+a^2(\eta){\cal M}^2(\eta)}} , 
\label{bulkdivleadDR}
\end{equation}
we can perform this loop integral, extracting the pole and the dependence on the renormalization scale $\mu$, 
\begin{equation}
- {\lambda\over 32\pi^2} \phi(\eta) {\cal M}^2(\eta) \left[ 
{1\over\epsilon} + 1 - \gamma + \ln{4\pi\mu^2\over a^2{\cal M}^2} \right] . 
\label{bulkdivpole}
\end{equation}
Adding this result to the other boundary-independent terms that appear in the tadpole integrand in Eq.~(\ref{leadingloops}) and substituting in the expression for the effective mass ${\cal M}$ from Eq.~(\ref{newcalMdef}) produces the following bulk effect corrected to first order, 
\begin{eqnarray}
&&\!\!\!\!\!\!\!\!\!\!\!\!\!\!\!\!\!\!
\langle \alpha_k(\eta_f) | \psi^+(x) | \alpha_k(\eta_f) \rangle
\nonumber \\
&=&
- i \int_{\eta_0}^{\eta_f} d\eta\, a^4(\eta) 
\left[ U_0^E(\eta_f) U_0^{E*}(\eta) - U_0^{E*}(\eta_f) U_0^E(\eta) 
\right] 
\nonumber \\
&&\quad\times\biggl\{
\nabla^2\phi 
+ m^2 \phi \left[ 
1 - {\lambda\over 32\pi^2} \left[ 
{1\over\epsilon} + 1 - \gamma + \ln{4\pi\mu^2\over{\cal M}^2} \right] 
\right]
\nonumber \\
&&\qquad
+ {\lambda\over 6} \phi^3 \left[ 
1 - {3\lambda\over 32\pi^2} \left[ 
{1\over\epsilon} + 1 - \gamma + \ln{4\pi\mu^2\over{\cal M}^2} \right] 
\right]
\nonumber \\
&&\qquad
+ R \phi \left[ \xi
- {\lambda\over 32\pi^2} 
\left( \xi - {1\over 6} \right) \left[ 
{1\over\epsilon} + 1 - \gamma + \ln{4\pi\mu^2\over{\cal M}^2} \right] 
\right]
\nonumber \\
&&\qquad
+ \cdots \biggr\} . 
\label{bulkrenorm}
\end{eqnarray}
In this expression we have omitted finite contributions which do not affect the renormalization or the running and we have also not written any of the terms that depend explicitly on the choice of the initial state, since these effects are discussed separately in the next section.  To fix the scale-independent part of the rescalings from the bare to the renormalized parameters, we apply the standard $\overline{\rm MS}$ prescription which sets 
\begin{equation}
Z_0 = 1 + {\lambda\over 32\pi^2} \Bigl[ 
{1\over\epsilon} + 1 - \gamma + \ln 4\pi \Bigr] + \cdots , 
\label{Z0oneloop}
\end{equation}
\begin{equation}
Z_1 = 1 + {3\lambda\over 32\pi^2} \Bigl[ 
{1\over\epsilon} - \gamma + \ln 4\pi \Bigr] + \cdots
\label{Z1oneloop}
\end{equation}
and
\begin{equation}
Z_\xi = 1 - {1\over\xi} \Bigl( \xi + {1\over 6} \Bigr) 
{\lambda\over 32\pi^2} \Bigl[ {1\over\epsilon} - \gamma + \ln 4\pi \Bigr] + \cdots
\label{Zxioneloop}
\end{equation}
to leading nontrivial order in the $\lambda$.  In terms of the renormalized parameters then, the tadpole is given by 
\begin{eqnarray}
&&\!\!\!\!\!\!\!\!\!\!\!\!\!\!\!\!\!\!
\langle \alpha_k^R(\eta_f) | \psi^+_R(x) | \alpha_k^R(\eta_f) \rangle
\nonumber \\
&=&
- i \int_{\eta_0}^{\eta_f} d\eta\, a^4(\eta) 
\left[ U_0^E(\eta_f) U_0^{E*}(\eta) - U_0^{E*}(\eta_f) U_0^E(\eta) 
\right] 
\nonumber \\
&&\quad\times\Bigl\{
\nabla^2\phi(\eta) 
+ m^2_R \phi(\eta) 
\Bigl[ 1 - {\lambda_R\over 16\pi^2} \ln{\mu\over{\cal M}_R(\eta)} \Bigr]
\nonumber \\
&&\qquad
+ {\lambda_R\over 6} \phi^3(\eta) 
\Bigl[ 1 - {3\lambda_R\over 16\pi^2} \ln{\mu\over{\cal M}_R(\eta)} \Bigr]
\nonumber \\
&&\qquad
+ R(\eta) \phi(\eta) \Bigl[ \xi_R - \Bigl( \xi_R - {1\over 6} \Bigr) 
{\lambda_R\over 16\pi^2} \ln{\mu\over{\cal M}_R(\eta)} \Bigr]
\nonumber \\
&&\qquad
+ \cdots \Bigr\} 
\label{bulkrenormR}
\end{eqnarray}
again up to finite and boundary-dependent contributions.

Despite the added complexity inherent in an inflationary environment, the cancellation of the bulk divergences has proceeded exactly as in a Minkowski space $S$-matrix calculation.  The reason for this simplicity is that although both the background and the initial state are nontrivial, neither has any effect on the bulk renormalization.  The divergences occur at infinitesimally short distances where the background curvature is not apparent, except as a classical source such as in the renormalization of the coupling of the field to the curvature.  Further, once we are sufficiently far from the boundary, how the field propagates through the bulk should be insensitive to the short-distance structure of that boundary.  Therefore, although there can be new divergences associated with the initial state, they are completely disjoint from the bulk divergences.

The renormalization scale $\mu$ is an artifact of our ignorance of the true, bare theory.  If we had calculated a matrix element solely in terms of the properties of the bare theory, the result would be completely independent of this scale, 
\begin{equation}
\mu {d\over d\mu} \langle\alpha_k(\eta_f) | \psi(x) | \alpha_k(\eta_f)\rangle 
= 0 . 
\label{callanbulk}
\end{equation}
But as the matrix element calculated in either the bare or the renormalized theory is exactly the same, up to a factor of the wave function rescaling, 
\begin{equation}
\langle\alpha_k(\eta_f) | \psi(x) | \alpha_k(\eta_f)\rangle 
= Z_3^{1/2}\langle\alpha_k^R(\eta_f) | \psi_R(x) | \alpha_k^R(\eta_f)\rangle  
\label{simmatrix}
\end{equation}
we have a similar equation for the scale dependence of the matrix element of the renormalized theory, 
\begin{equation}
\left[ \mu {d\over d\mu} + \gamma(\lambda_R) \right] 
\langle\alpha_k^R(\eta_f) | \psi_R(x) | \alpha_k^R(\eta_f)\rangle = 0 , 
\label{callanbulkZ}
\end{equation}
where $\gamma(\lambda_R)$ is the anomalous dimension of the field, 
\begin{equation}
\gamma(\lambda_R) = {1\over 2} \mu {d\over d\mu} \ln Z_3 . 
\label{anomdef}
\end{equation}
All the renormalized parameters depend upon the renormalization scale so we introduce the usual functions that describe their running,
\begin{equation}
\beta(\lambda_R) = \mu {d\lambda_R\over d\mu}, 
\quad\ 
\gamma_m(\lambda_R) = {\mu\over m_R} {dm_R\over d\mu}, 
\quad\ 
\beta_\xi(\lambda_R) = \mu {d\xi_R\over d\mu}, 
\label{runnings}
\end{equation}
in terms of which we obtain the Callan-Symanzik equation, 
\begin{eqnarray}
&&\!\!\!\!\!
\biggl[ 
\mu {\partial\over\partial\mu} 
+ \beta(\lambda_R) {\partial\over\partial\lambda_R} 
+ m_R \gamma_m(\lambda_R) {\partial\over\partial m_R} 
\nonumber \\
&&\!\!\!\!\!
+ \beta_\xi(\lambda_R) {\partial\over\partial\xi_R} 
+ \gamma(\lambda_R)
+ \cdots 
\biggr] \langle\alpha_k^R(\eta_f) | \psi(x) | \alpha_k^R(\eta_f)\rangle 
= 0 
\nonumber \\
&&
\label{callanbulkR}
\end{eqnarray}
up to corrections which do depend on the boundary conditions.  From the expression for the renormalized tadpole, we have the standard running of the bulk parameters at leading order, 
\begin{eqnarray}
\beta(\lambda_R) &=& {3\lambda_R\over 16\pi^2} + \cdots , 
\nonumber \\
\gamma_m(\lambda_R) &=& {\lambda_R\over 32\pi^2} + \cdots , 
\nonumber \\
\beta_\xi(\lambda_R) &=& \Big[ \xi_R - {1\over 6} \Bigr] 
{\lambda_R\over 16\pi^2} + \cdots . 
\label{betafunc}
\end{eqnarray}
Notice that the coupling to the curvature does not run at this order when the field is conformally coupled, i.e.\ $\xi_R={1\over 6}$. 

The Callan-Symanzik equation depends upon the running of both the bulk parameters and those describing the boundary since once we have imposed a renormalization condition there exists a single renormalization scale $\mu$ in the theory.  The fact that there is a unique scale for both, rather than separate scales, is particularly clear from a Wilsonian perspective \cite{wilson}.  In the functional integral description of the theory \cite{greens}, the generating functional contains information about the bulk theory, in the action, and the initial state, in the source term.  To understand how the short-distance physics affects what we mean by a particular parameter, whether for the bulk physics or the initial state, we start with theory defined up to a cutoff, $\Lambda$, which is equivalent to truncating the functional integral for modes above this scale.  We can then determine how the parameters flow as we alter the cutoff scale from $\Lambda$ to $\Lambda'$ $(< \Lambda)$ by integrating out the field modes between these scales.  The resulting effective theory based on the cutoff $\Lambda'$ will have its bulk parameters affected in the usual way.  But the form of our effective description of the initial state will also be shifted as well, since how accurately we can describe the features of the initial state also depends on the field modes available in our functional integral.  Thus only one quantity, whether $\Lambda'/\Lambda$ for a cutoff theory or $\mu$ for a dimensionally regularized theory, is needed to describe the running of both the bulk and boundary effects since the origin of this running is common to both aspects of theory.

\section{Boundary renormalization}
\label{boundary}

Thus far, we have written the structure function for the initial state, $e^{\alpha_k^*}$, without specifying how it varies with the momentum.  We shall require a more detailed form to show under what conditions a general initial state, when renormalized, produces a theory with finite perturbative corrections.  If this renormalization is genuinely associated only with the details of the initial state, then it should be sufficient to apply the renormalization only at the initial time and the theory should remain finite at all subsequent times.  This reasoning suggests that if we regard the rescaling of the initial state as the appropriate addition of counterterms, chosen to cancel the divergences at the initial time, then these counterterms should be expressed through a three dimensional boundary action, 
\begin{equation}
S_{\eta=\eta_0} = \int d^3\vec x\, \sqrt{-h} {\cal L}_{3\rm d}(\varphi) . 
\label{bdnaction}
\end{equation}
$h$ denotes the determinant of the induced metric $h_{\mu\nu}$ on the $\eta=\eta_0$ boundary.  Using again the time-like normal vector $n^\mu$ that is orthogonal to the boundary, the induced metric is obtained from the full metric by 
\begin{equation}
h_{\mu\nu} = g_{\mu\nu} - n_\mu n_\nu . 
\label{induced}
\end{equation}
For the conformally flat metric, 
\begin{equation}
h_{\mu\nu}\, dx^\mu dx^\nu = - a^2(\eta)\, d\vec x\cdot d\vec x 
\label{confh}
\end{equation}
and $\sqrt{-h} = a^3(\eta)$.  From the induced metric we can construct an induced scalar curvature, which we denote by $\hat R$, while the extrinsic curvature tensor is defined by projecting the covariant derivative of the normal back onto the initial surface, 
\begin{equation}
K_{\mu\nu} = h_\mu^{\ \lambda} \nabla_\lambda n_\nu . 
\label{extrinsic}
\end{equation}
This tensor as well as its trace, $K = h^{\mu\nu} K_{\mu\nu}$, provide additional dimension 1 ingredients out of which we can construct operators for the boundary action.  For an expanding background, it is proportional to the Hubble parameter,
\begin{equation}
K(\eta) = {3H(\eta)\over a(\eta)} . 
\label{extrinsicRW}
\end{equation}

The operators contained in the counterterm Lagrangian are classified according to whether their mass dimension is greater or less than, or equal to, the dimension of the boundary, but the fields inherit their scaling dimension from the full $3+1$ dimensional theory.  Thus, for example, in our scalar theory with a $\varphi \leftrightarrow - \varphi$ symmetry, there is a unique relevant operator of dimension 2, 
\begin{equation}
\varphi^2 , 
\label{relevant}
\end{equation}
while there are two marginal operators, 
\begin{equation}
\varphi\nabla_n\varphi ,\quad K\varphi^2 . 
\label{marginal}
\end{equation}
Compared with a flat background, the number of higher dimensional operators rapidly proliferates since we can also build counterterms with curvature-dependent objects such as the extrinsic curvature or the induced curvature.  For example, the complete set of irrelevant, dimension 4 operators consists of 
\begin{eqnarray}
&&\varphi^4, \varphi\nabla_n^2\varphi, (\nabla_n\varphi)^2, 
\vec\nabla\varphi \cdot \vec\nabla\varphi, 
\nonumber \\
&&
K^2\varphi^2, K^{\mu\nu}K_{\mu\nu}\varphi^2, 
K\varphi\nabla_n\varphi, (\nabla_n K)\varphi^2, \hat R\varphi^2 . 
\label{irrelevant}
\end{eqnarray}
For particular, highly symmetric backgrounds, only a subset of this list might be required.  Only the first four are needed in flat space and the complete list for de Sitter space is 
\begin{equation}
\varphi^4, \varphi\nabla_n^2\varphi, (\nabla_n\varphi)^2, 
\vec\nabla\varphi \cdot \vec\nabla\varphi, 
K^2\varphi^2, K\varphi\nabla_n\varphi . 
\label{dSdim4}
\end{equation}

The idea then is to label specific aspects of the initial state by the type of counterterms associated with them so that a particular piece of the boundary condition is relevant, marginal or irrelevant according to the dimension of the boundary counterterms needed to remove the divergences it produces.  Alternatively, a boundary condition can also be characterized as renormalizable or nonrenormalizable according to whether its difference from the vacuum state diminishes or grows at shorter and shorter distances.  The nonrenormalizable initial states are not necessarily nonpredictive---they can be understood as an effective description of the state.  For example, if we associate a heavy mass scale $M$ with these effects, it generically will require an infinite set of dimension $n>3$ counterterms to render the theory finite.  However, since these terms are suppressed by factors of $(H/M)^{n-3}$, as long as $H/M\ll 1$ only a very small subset of the parameters describing a general nonrenormalizable initial state are required in practice.

We shall describe a general initial state by an expansion in the generalized frequency, $\Omega_k(t_0)$, evaluated at the initial surface, 
\begin{equation}
e^{\alpha_k^*} = \sum_{n=0}^\infty d_n {H^n(\eta_0)\over\Omega_k^n(\eta_0)} 
+ \sum_{n=1}^\infty c_n {\Omega_k^n(\eta_0)\over a^n(\eta_0)M^n} . 
\label{ealphser}
\end{equation}
These two series are respectively associated with the infrared and ultraviolet aspects of the initial state.  We have chosen an expansion in the frequency, rather than just the momentum $k$, since this quantity has several useful properties.  It is finite, although of a complicated form, in the $k\to 0$ limit and it has a natural transition in its behavior at $k\sim H(\eta_0)$.  Above this scale $\Omega_k(\eta_0)\to k$, while below this scale the curvature-dependent effects become dominant compared with its explicit momentum dependence.

In the first series, we need a reasonable dynamical scale for separating the long from the short distances on our initial surface so we have chosen the expansion rate, $H(\eta_0)$, to set this scale.  Since at extremely short distances, where $\Omega_k^n(\eta_0)\sim k^n$, these terms become diminishingly important once $k\gg H(\eta_0)$.  More generally, we should use a linear combination of $m$ and $H(\eta_0)$ in the numerators of this expansion to obtain the a quantity which does not vanish entirely in the Minkowski space limit, but in slowly rolling models of inflation $m\ll H(\eta)$ so we shall neglect the mass in this expansion.  Notice that in the extreme ultraviolet limit, $k\to\infty$, all of the terms in this series vanish except for the marginal term, $d_0$.  

Thus the first series describes features of the state which are important at long distances and can be viewed as some nonvacuum ensemble since in this regime the idea of the vacuum defined with respect to the free Klein-Gordon equation holds very well.  The signals of trans-Planckian physics lies properly in the second series in Eq.~(\ref{ealphser}).  Unlike the terms in the first series, these grow in importance at extremely short distances.  Assuming that the scale of new physics is above the Hubble scale, $M\gg H(\eta_0)/a(\eta_0)$, these terms are essentially a series in $k/M$.  Since the structure function $e^{\alpha_k^*}$ accompanies the initial state contribution to the propagator and to the modes, it might seem that these growing terms do not describe a sensible state at extremely short distances.  But if we are measuring some process at a scale much lower than $M$, the contributions from the $n^{\rm th}$ term in this series is naturally suppressed by the $n^{\rm th}$ power of the ratio of the scale we measure to $M$.  Of course, loop corrections sample over all momenta, so the large difference between the state and the extrapolated vacuum will produce divergences in these corrections; but since these divergences occur from the large momentum---short distance---part of the loop integral evaluated at the initial time, they can always be cancelled by adding the appropriate local counterterms on the initial boundary.

In \cite{greens} we showed that the divergences from the new initial condition only appear at the initial-time surface, $\eta=\eta_0$.  More precisely, the divergences only occur in a boundary loop correction when the term is {\it simultaneously\/} evaluated at $\eta=\eta_0$ {\it and\/} we sum over arbitrarily large values of the loop momenta.  Since the Bunch-Davies state matches with the flat space vacuum states at large values of the momenta, the theory in a curved background inherits exactly the same divergence structure from the short-distance features of the initial state as flat space.  The only new feature is that the invariants associated with the curvature allow for a richer family of boundary counterterms, as was seen in Eq.~(\ref{irrelevant}).  

The prescription for extracting the short-distance boundary divergences is then as follows.  We first define a family of kernel functions, 
\begin{equation}
K^{(p)}(\eta) \equiv 
\int {d^3\vec k\over (2\pi)^2}\, 
{e^{-2i\int_{\eta_0}^\eta d\eta'\, \Omega_k(\eta')}
\over \Omega_k^{3-p}(\eta) \Omega_k^p(\eta_0)} , 
\label{Kpkerneldef}
\end{equation}
which arise from inserting the expansion of the initial state structure function in Eq.~(\ref{ealphser}) into a loop integral.  In this kernel, the terms evaluated at $\eta_0$ are associated with the series expansion for the initial state and the remaining $\eta$-dependent factors are associated with the loop propagators.  These kernel functions have been constructed so that they only contain a mild---integrable---logarithmic singularity at $\eta=\eta_0$.  The loop integrals do not always appear exactly in the form of one of these kernels, but they can always be expressed in terms of some $n^{\rm th}$ derivative of them, up to explicitly finite terms.  By then integrating these kernel-derivatives by parts $n$ times, we obtain a finite term where $K^{(p)}(\eta)$ appears in the $d\eta$-integrand as well as a set of boundary terms, some evaluated at $\eta_f$ and some at $\eta_0$.  The former are finite while the latter are divergent and can be regularized using dimensional regularization.  The resulting set of divergent terms determines the set of boundary counterterms that we should add to the theory to render it finite.  The behavior of these kernels, their regularization and their derivatives, is explained more fully in Appendix~\ref{kernels}.

We begin in the next subsection first with an illustration of this procedure for the renormalizable initial state.  Although the prescription applies to any type of initial state, it is simpler in the case of the renormalizable state since fewer integrations by parts are necessary.  What we shall find is that for a theory with a $\varphi^4$ interaction, only the first two terms ($d_0$ and $d_1$) require any renormalization and are associated with the renormalizable set of counterterms, $\{ \varphi^2, \varphi\nabla_n\varphi, K\varphi^2 \}$.

\subsection{Renormalizable boundary conditions}

Renormalizable boundary conditions correspond to those which may differ substantially from the Bunch-Davies vacuum at long distances but which become indistinguishable from this vacuum at very short distances.  These states therefore do not generally contain information about trans-Planckian physics and resemble some excited ensemble, from the perspective of the low energy, weakly interacting theory.  Despite this simplicity at short distances, in a few cases where the difference between the initial and the Bunch-Davies states does not diminish sufficiently fast, these initial states can produce new divergences isolated at the initial time.  The renormalizable states thus provide the simplest example of the boundary renormalization which is necessary for a general initial state and which results in a controlled, finite perturbative description of the interacting theory.

If we restrict to an initial condition described only by negative powers of the generalized frequency, $\Omega_k(\eta_0)$, 
Eq.~(\ref{ealphser}), 
\begin{equation}
e^{\alpha_k^*} = \sum_{n=0}^\infty d_n {H^n(\eta_0)\over\Omega_k^n(\eta_0)} , 
\label{IRbound}
\end{equation}
we can obtain a sense of which terms will require renormalization at the initial boundary by expanding the initial condition on the modes in Eq.~(\ref{incond}) at $\eta=\eta_0$ and looking at the short distance, $k\to\infty$, limit,
\begin{eqnarray}
{1\over a(\eta_0)} {\partial\over\partial\eta} U_k(\eta_0) 
&=& -i \varpi_k U_k(\eta_0) 
\nonumber \\
&\approx& 
-ik U_k(\eta_0) 
\nonumber \\
&&
+ i \left[ {2d_0\over 1+d_0} k + {2H(\eta_0)d_1\over (1+d_0)^2} 
\right] U_k(\eta_0) . 
\nonumber \\
&& 
+ {\cal O}\left( {H^2(\eta_0)\over k} \right) U_k(\eta_0) . 
\label{IRbndlimit}
\end{eqnarray}
At momenta well above the Hubble scale, $k\gg H(\eta_0)$, all the terms aside from $d_0$ and $d_1$ have no effect.

The leading contribution to the tadpole that depends on the details of the initial state appears in the last term included in Eq.~(\ref{leadingloops}), 
\begin{equation}
- {\lambda\over 2} \int_{\eta_0}^{\eta_f} dt\, 
a^4(\eta) {\cal G}(\eta_f,\eta) \phi(\eta) 
\int {d^3\vec k\over (2\pi)^3}\, e^{\alpha_k^*} U_k^E(\eta) U_k^E(\eta) . 
\label{leadbndone}
\end{equation}
In terms of the adiabatic modes and the leading terms of our expansion for $e^{\alpha_k^*}$ given in Eq.~(\ref{IRbound}), this contribution becomes 
\begin{widetext}
\begin{eqnarray}
&&\!\!\!\!\!\!\!\!\!\!\!\!\!\!\!\!\!\!\!\!\!\!\!
- {\lambda\over 2} \int_{\eta_0}^{\eta_f} dt\, 
a^4(\eta) {\cal G}(\eta_f,\eta) \phi(\eta) 
\int {d^3\vec k\over (2\pi)^3}\, e^{\alpha_k^*} U_k^E(\eta) U_k^E(\eta)
\nonumber \\
&=&
- {\lambda\over 4} 
\int_{\eta_0}^{\eta_f} d\eta\, a^2(\eta) {\cal G}(\eta_f,\eta) \phi(\eta)
\biggl\{
d_0 \int {d^3\vec k\over (2\pi)^3}\, 
{e^{-2i\int_{\eta_0}^\eta d\eta'\, \Omega_k(\eta')}\over \Omega_k(\eta)} 
+ d_1 H(\eta_0) 
\int {d^3\vec k\over (2\pi)^3}\, 
{e^{-2i\int_{\eta_0}^\eta d\eta'\, \Omega_k(\eta')}\over \Omega_k(\eta) \Omega_k(\eta_0)} 
\nonumber \\
&&\qquad\qquad\qquad\qquad\qquad\qquad\quad
+\ d_2 H^2(\eta_0) 
\int {d^3\vec k\over (2\pi)^3}\, 
{e^{-2i\int_{\eta_0}^\eta d\eta'\, \Omega_k(\eta')}\over \Omega_k(\eta) \Omega_k^2(\eta_0)} 
+ \cdots \biggr\} .  
\label{leadbndoneAD}
\end{eqnarray}
\end{widetext}
The terms associated with higher moments are associated with loop integrals that are manifestly finite.  For the $n^{\rm th}$ term in the moment expansion, the large momentum region of the accompanying loop integral behaves as 
\begin{eqnarray}
d_n \int {d^3\vec k\over (2\pi)^3}\, 
{e^{-2i\int_{\eta_0}^\eta d\eta'\, \Omega_k(\eta')}\over 
\Omega_k(\eta)\Omega_k^n(\eta_0)}
&=& 
{d_n\over 2\pi^2} \int_{k\to\infty} {dk\over k^{n-1}}\, 
e^{-2ik(\eta-\eta_0)} 
\nonumber \\
&&
+ \cdots
\label{UVfinite}
\end{eqnarray}
which is finite for $n>2$.  The cases $n=0,1,2$ produce a quadratic pole, a simple pole and a logarithmic singularity at $\eta=\eta_0$, respectively \cite{greens}.  Because of the remaining conformal time integration in Eq.~(\ref{leadbndoneAD}), the logarithmic singularity actually only gives a finite contribution to the tadpole so that only the first two terms produce divergences, as was suggested by examining the short-distance structure of the initial condition.  Therefore, for the rest of this section we shall set $d_n=0$ for $n\ge 2$.

To show that the divergences associated with the initial state are genuinely confined to the initial surface, let us reexpress the loop integrals as derivatives of the kernels defined earlier in Eq.~(\ref{Kpkerneldef}).  Just as in the case of the $d_2$ term, these kernels are only logarithmically divergent in $(\eta-\eta_0)$, so that with an appropriate number of integrations by parts we can isolate the divergent pieces explicitly.  Noting that 
\begin{equation}
\int {d^3\vec k\over (2\pi)^3}\, 
{e^{-2i\int_{\eta_0}^\eta d\eta'\, \Omega_k(\eta')}\over \Omega_k(\eta) }
= - {1\over 4} K^{(0)\, \prime\prime}(\eta) + \hbox{UV finite}
\label{kerneldiffRa} 
\end{equation}
and
\begin{equation}
\int {d^3\vec k\over (2\pi)^3}\, 
{e^{-2i\int_{\eta_0}^\eta d\eta'\, \Omega_k(\eta')}
\over \Omega_k(\eta) \Omega_k(\eta_0)}
= - {1\over 2i} K^{(1)\, \prime}(\eta) 
+ \hbox{UV finite}
\label{kerneldiffRb} 
\end{equation}
and substituting these expressions into Eq.~(\ref{leadbndoneAD}) yields,
\begin{widetext}
\begin{eqnarray}
&&\!\!\!\!\!\!\!\!\!\!\!\!\!\!\!\!\!\!\!\!\!\!\!
- {\lambda\over 2} \int_{\eta_0}^{\eta_f} dt\, 
a^4(\eta) {\cal G}(\eta_f,\eta) \phi(\eta) 
\int {d^3\vec k\over (2\pi)^3}\, e^{\alpha_k^*} U_k^E(\eta) U_k^E(\eta)
\nonumber \\
&=&
{\lambda d_0\over 16} \biggl\{
- a^2(\eta_0) {\cal G}(\eta_f,\eta_0) \phi(\eta_0)
K^{(0)\, \prime}(\eta_0)
- a^2(\eta_f) \partial_\eta {\cal G}(\eta_f,\eta)\bigr|_{\eta=\eta_f} \phi(\eta_f) 
K^{(0)}(\eta_f)
\nonumber \\
&&\qquad
+ \partial_\eta \bigl[ a^2(\eta) {\cal G}(\eta_f,\eta) \phi(\eta) \bigr]_{\eta=\eta_0}
K^{(0)}(\eta_0)
+ \int_{\eta_0}^{\eta_f} d\eta\, 
\partial^2_\eta \bigl[ a^2(\eta) {\cal G}(\eta_f,\eta) \phi(\eta) \bigr]
K^{(0)}(\eta)
\biggr\}
\nonumber \\
&&
- {\lambda d_1\over 8i} H(\eta_0) \biggl\{
a^2(\eta_0) {\cal G}(\eta_f,\eta_0) \phi(\eta_0) K^{(1)}(\eta_0)  
+ \int_{\eta_0}^{\eta_f} d\eta\, 
\partial_\eta \bigl[ a^2(\eta) {\cal G}(\eta_f,\eta) \phi(\eta) \bigr]
K^{(1)}(\eta) 
\biggr\}
+ \cdots 
\nonumber \\
&=&
{\lambda d_0\over 16} 
\partial_\eta \bigl[ a^2(\eta) {\cal G}(\eta_f,\eta) \phi(\eta) \bigr]_{\eta=\eta_0}
K^{(0)}(\eta_0)
- {\lambda d_1\over 8i} H(\eta_0) 
a^2(\eta_0) {\cal G}(\eta_f,\eta_0) \phi(\eta_0) K^{(1)}(\eta_0) 
+ {\rm finite}
\label{leadbndoneADK}
\end{eqnarray}
after integrating by parts and using that ${\cal G}(\eta_f,\eta_f) = 0$.  The singular kernels can be regularized by extending the number of spatial dimensions to $3-2\epsilon$ as has been done in Appendix~\ref{kernels} which allows us to extract the poles as $\epsilon\to 0$ as well as the dependence on the renormalization scale $\mu$, 
\begin{eqnarray}
&&\!\!\!\!\!\!\!\!\!\!\!\!\!\!\!\!\!\!\!\!\!\!\!
- {\lambda\over 2} \int_{\eta_0}^{\eta_f} dt\, 
a^4(\eta) {\cal G}(\eta_f,\eta) \phi(\eta) 
\int {d^3\vec k\over (2\pi)^3}\, e^{\alpha_k^*} U_k^E(\eta) U_k^E(\eta)
\nonumber \\
&=&
{\lambda d_0\over 64\pi^2} 
\partial_\eta \bigl[ a^2(\eta) {\cal G}(\eta_f,\eta) \phi(\eta) \bigr]_{\eta=\eta_0}
\left[ {1\over\epsilon} - \gamma + \ln{4\pi\mu^2\over a^2(\eta_0){\cal M}^2(\eta_0)} \right]
\nonumber \\
&&
- {\lambda d_1\over 32i\pi^2} H(\eta_0) 
a^2(\eta_0) {\cal G}(\eta_f,\eta_0) \phi(\eta_0) 
\left[ {1\over\epsilon} - \gamma + \ln{4\pi\mu^2\over a^2(\eta_0){\cal M}^2(\eta_0)} \right]
+ {\rm finite} . 
\label{leadbndoneDIV}
\end{eqnarray}
\end{widetext}

These are the divergences associated with the ``bare'' initial condition and can be renormalized by adding the following counterterms to the interaction Hamiltonian 
\begin{eqnarray}
H_I^{\rm c.t.} 
&=& \int d^3\vec y\, a^4(\eta) \biggl\{ 
{z_0\over 2} {\delta'(\eta-\eta_0)\over a^2(\eta)}\, \phi\psi^\pm
\nonumber \\
&&\qquad\qquad\quad 
+ {z_1\over 2} {\delta(\eta-\eta_0)\over a(\eta)} K\phi\psi^\pm
\biggr\} \qquad
\label{bndham}
\end{eqnarray}
which vanish except at the initial boundary.  We can equivalently regard these terms as the addition of the following three-dimensional boundary action to the theory, 
\begin{equation}
S_{\rm 3d} 
= {\textstyle{1\over 2}} \int_{\eta_0} d^3\vec y\, \sqrt{-h} 
\left\{ 
z_0 n^\mu \nabla_\mu \bigl( \phi\psi^\pm \bigr)
- \left( z_1 - {\textstyle{2\over 3}} z_0 \right) K\phi\psi^\pm
\right\} . 
\label{bndaction}
\end{equation}
Note that we could have also included the dimension 2 counterterm, $m\phi\psi^\pm$, but its role is suppressed in an inflationary background where the extrinsic curvature term provides the dominant contribution.  The leading contribution to the tadpole from these counterterms is 
\begin{eqnarray}
&&\!\!\!\!\!\!\!\!\!\!\!\!\!\!\!\!\!\!\!\!\!
\langle\alpha_k^R(t_f)|\psi_R^+(x)|\alpha_k^R(t_f)\rangle 
\nonumber \\
&=& 
\cdots + {\textstyle{1\over 2}} z_0 
\partial_\eta \bigl[ a^2(\eta) {\cal G}(\eta_f,\eta) \phi(\eta) \bigr]_{\eta_0}
\nonumber \\
&&
-\ {\textstyle{1\over 2}} z_1 a^3(\eta_0) {\cal G}(\eta_f,\eta_0) \phi(\eta_0) K(\eta_0) . 
\label{tadct}
\end{eqnarray}

The inclusion of the boundary counterterms is the boundary analogue of scaling the bulk parameters which translates between the bare and the renormalized theories.  Because of this role, the counterterms will properly contain two components, 
\begin{eqnarray}
z_0 &=& z_0^\epsilon + \hat z_0(\mu) 
\nonumber \\
z_1 &=& z_1^\epsilon + \hat z_1(\mu) , 
\label{littlezs}
\end{eqnarray}
representing the infinite scale-independent piece, which is fixed by the renormalization scheme, and a finite scale-dependent piece, which is fixed by the overall scale independence of the matrix element.  In the $\overline{\rm MS}$ scheme, the scale-independent part is fixed to cancel the pole and the usual finite artifacts of dimensional regularization,
\begin{eqnarray}
z_0^\epsilon 
&=& - {\lambda\over 32\pi^2} d_0 
\Bigl[ {1\over\epsilon} - \gamma + \ln 4\pi \Bigr]
\nonumber \\
z_1^\epsilon 
&=& - {\lambda\over 48i\pi^2} d_1
\Bigl[ {1\over\epsilon} - \gamma + \ln 4\pi \Bigr] . 
\label{MSbarbnd}
\end{eqnarray}
The resulting contribution to the tadpole from those parts of the renormalized initial condition that depend on the renormalization scale is contained in the terms, 
\begin{eqnarray}
&&\!\!\!\!\!\!\!\!\!\!\!\!\!\!\!\!\!\!
\langle\alpha_k^R(\eta_f)|\psi_R^+(x)|\alpha_k^R(\eta_f)\rangle
\nonumber \\
&=&
{1\over 2} \left[ 
\hat z_0(\mu)
+ 
{\lambda_R d_0\over 16\pi^2} 
\ln{\mu\over a{\cal M}_R} 
\right]
\partial_\eta \bigl[ a^2 {\cal G}(\eta_f,\eta) \phi \bigr]_{\eta=\eta_0}
\nonumber \\
&&
- {1\over 2} \left[ 
\hat z_1(\mu) 
+ 
{\lambda_R d_1\over 24i\pi^2} 
\ln{\mu\over a{\cal M}_R} 
\right] 
a^3 {\cal G}(\eta_f,\eta_0) K\phi \bigr|_{\eta=\eta_0} 
\nonumber \\
&&
+ \cdots . 
\label{renbnd}
\end{eqnarray}

Among the many terms not explicitly written in this equation are all the other boundary-independent, $\mu$-dependent contributions from the bulk, evaluated in Eq.~(\ref{bulkrenormR}).  If we introduce boundary $\beta$-functions for $\hat z_0$ and $\hat z_1$,
\begin{eqnarray}
\hat\beta_0(\lambda_R) &=& \mu {d\hat z_0\over d\mu} 
\nonumber \\
\hat\beta_1(\lambda_R) &=& \mu {d\hat z_1\over d\mu} , 
\label{hatbetas}
\end{eqnarray}
then from Eq.~(\ref{renbnd}) we have that 
\begin{eqnarray}
\hat\beta_0(\lambda_R) &=& - {\lambda_R d_0\over 16\pi^2} + \cdots 
\nonumber \\
\hat\beta_1(\lambda_R) &=& - {\lambda_R d_1\over 24i\pi^2} + \cdots . 
\label{hatbetavals}
\end{eqnarray}
Here we have used the Callan-Symanzik equation 
\begin{eqnarray}
&&\!\!\!\!\!
\biggl[ 
\mu {\partial\over\partial\mu} 
+ \beta(\lambda_R) {\partial\over\partial\lambda_R} 
+ m_R \gamma_m(\lambda_R) {\partial\over\partial m_R} 
\nonumber \\
&&\!\!\!\!\!
+ \beta_\xi(\lambda_R) {\partial\over\partial\xi_R} 
+ \gamma(\lambda_R)
+ \hat\beta_0(\lambda_R) {\partial\over\partial\hat z_0} 
+ \hat\beta_1(\lambda_R) {\partial\over\partial\hat z_1} 
\nonumber \\
&&\!\!\!\!\!
+ \cdots 
\biggr] \langle\alpha_k^R(\eta_f) | \psi_R^+(x) | \alpha_k^R(\eta_f)\rangle 
= 0 , 
\label{callanbulkRR}
\end{eqnarray}
which now includes the renormalizable initial condition parameters as well as the bulk parameters, to determine the running of the boundary conditions.  The equation is not yet in its complete form since we have not included the nonrenormalizable initial effects, which we discuss next.

\subsection{Nonrenormalizable boundary conditions}

The signals of trans-Planckian physics reside in the nonrenormalizable part of the initial state.  Such an initial state is one which differs increasingly from the the vacuum state at shorter distances or which equivalently requires nonrenormalizable counterterms in the boundary action to render the theory finite.  In terms of our expansion, these features of the initial state are those described by a series of positive powers of the generalized frequency, 
\begin{equation}
e^{\alpha_k^*} = \sum_{n=1}^\infty c_n 
{\Omega_k^n(\eta_0)\over a^n(\eta_0) M^n} . 
\label{UVseries}
\end{equation}
$M$, as usual, represents the scale at which some new dynamics becomes important.  

A great advantage of an effective description of an initial state is in its applicability to any idea for how the physics above the expansion scale might be modified.  Different ideas---minimum lengths, modified uncertainty relations or new dispersion relations---can be distinguished by the values of their coefficients $c_n$ in the series of Eq.~(\ref{UVseries}).  Even more importantly for observations, since at best only the leading nonvanishing term is likely to be observable, the general trans-Planckian signal is determined by the single quantity $c_1/M$, assuming $c_1$ is the first nonvanishing coefficient.  This property allows us to extract a generic prediction for a trans-Planckian signal in the cosmic microwave background which can then be contrasted with other ways for generating departures from the simple vacuum prediction, such as through modifications of the inflaton potential.

Since our goal here is to show how renormalization proceeds for a nonrenormalizable state and that perturbative corrections remain small when $H(\eta_0)\ll M$, we shall examine the renormalization of a state for the leading trans-Planckian effect with
\begin{equation}
e^{\alpha_k^*} = c_1 {\Omega_k(\eta_0)\over a(\eta_0)M} \left[
1 + {1\over 2} {a^2(\eta_0) m^2\over\Omega_k^2(\eta_0)} 
\right] . 
\label{UVserieseg}
\end{equation}
Since $\lim_{k\to 0}\Omega_k(\eta_0) \not= 0$, the nonrenormalizable terms do not vanish at long distances so we have added an extra renormalizable term, scaling as $1/\Omega_k(\eta)$, to subtract some of this long distance behavior so that the condition represents a genuine trans-Planckian effect.

The one loop contribution from this boundary condition to the tadpole is of the form 
\begin{eqnarray}
&&\!\!\!\!\!\!\!\!\!\!\!\!\!
\langle \alpha_k(\eta_f) | \psi^+(x) | \alpha_k(\eta_f) \rangle 
\nonumber \\
&=& - {\lambda\over 4} {c_1\over a(\eta_0)M} 
\int_{\eta_0}^{\eta_f} d\eta\, a^2(\eta) {\cal G}(\eta_f,\eta) \phi(\eta) 
\nonumber \\
&&\times
\int {d^3\vec k\over (2\pi)^3}\, 
\left[
1 + {1\over 2} {a^2(\eta_0) m^2\over\Omega_k^2(\eta_0)} 
\right] 
{\Omega_k(\eta_0) e^{-2i\int_{\eta_0}^\eta d\eta'\, \Omega_k(\eta')} 
\over \Omega_k(\eta)}
\nonumber \\
&&
+ \cdots . 
\label{leadUV}
\end{eqnarray}
Introducing the kernel 
\begin{equation}
\int {d^3\vec k\over (2\pi)^3}\, 
{\Omega_k(\eta_0) e^{-2i\int_{\eta_0}^\eta d\eta'\, \Omega_k(\eta')}\over \Omega_k(\eta) }
= {1\over 8i} K^{(-1)\, \prime\prime\prime}(\eta) + \hbox{UV finite}
\label{kerneldiffRc} 
\end{equation}
to represent the divergent part of the first term in the loop integral and $K^{(1)\, \prime}(\eta)$ as before in Eq.~(\ref{kerneldiffRb}) for the second, we integrate by parts until all the divergent contributions appear explicitly as term evaluated on the boundary, 
\begin{eqnarray}
&&\!\!\!\!\!\!\!\!\!\!\!\!\!\!\!\!\!\!\!
\langle \alpha_k(\eta_f) | \psi^+(x) | \alpha_k(\eta_f) \rangle 
\nonumber \\
&=& {\lambda\over 32i} {c_1\over M} {1\over a(\eta_0)} 
\partial^2_\eta \left[ a^2(\eta) {\cal G}(\eta_f,\eta) \phi(\eta) \right]_{\eta_0}
K^{(-1)}(\eta_0)
\nonumber \\
&& 
+ {\lambda\over 32i} {c_1\over M} 
a(\eta_0) {\cal G}(\eta_f,\eta_0) \phi(\eta_0) 
\nonumber \\
&&\qquad\times
\left[
K^{(-1)\, \prime\prime}(\eta_0) 
- 2 a^2(\eta_0) m^2 K^{(1)}(\eta_0)
\right]
\nonumber \\
&&
+ \cdots . 
\label{leadbndoneADUV}
\end{eqnarray}
Here we have only retained the terms that would diverge if we performed the remaining loop integrations.  Using the expressions for dimensionally regularized kernels from Appendix~\ref{kernels}, we find that 
\begin{eqnarray}
&&\!\!\!\!\!\!\!\!\!\!\!\!\!\!\!\!\!\!\!
\langle \alpha_k(\eta_f) | \psi^+(x) | \alpha_k(\eta_f) \rangle 
\nonumber \\
&=& {\lambda\over 128i\pi^2} {c_1\over M} 
\left[ 
{1\over\epsilon} - \gamma + \ln{4\pi\mu^2\over a^2(\eta_0){\cal M}^2(\eta_0)} 
\right]
\nonumber \\
&&\qquad\times 
{1\over a(\eta_0)} \partial^2_\eta \left[ a^2(\eta) {\cal G}(\eta_f,\eta) \phi(\eta) \right]_{\eta=\eta_0}
\nonumber \\
&& 
+ {\lambda\over 64i\pi^2} {c_1\over M} 
\left[ 
{1\over\epsilon} + 1 - \gamma + \ln{4\pi\mu^2\over a^2(\eta_0){\cal M}^2(\eta_0)} 
\right]
\nonumber \\
&&\qquad\times
a^3(\eta_0) {\cal G}(\eta_f,\eta_0) \left[ \xi - {1\over 6} \right] \phi(\eta_0) R(\eta_0)
\nonumber \\
&&
+ {\lambda^2\over 128i\pi^2} {c_1\over M} 
\left[
{1\over\epsilon} + 1 - \gamma + \ln{4\pi\mu^2\over a^2(\eta_0){\cal M}^2(\eta_0)} 
\right]
\nonumber \\
&&\qquad\times
a^3(\eta_0) {\cal G}(\eta_f,\eta_0) \phi^3(\eta_0) 
\nonumber \\
&&
+ \cdots . 
\label{leadbndoneADUVD}
\end{eqnarray}

These divergences are cancelled by adding the following counterterms to the interaction Hamiltonian, 
\begin{eqnarray}
H_I^{\rm c.t.} &=& \int d^3\vec y\, {a^4(\eta)\over a(\eta_0)}\, \biggl\{ 
{1\over 2} {z_2\over M}\, {\delta^{\prime\prime}(\eta-\eta_0)\over a^2(\eta)} \phi\psi^\pm 
\nonumber \\
&&\qquad\qquad 
+ {z_3\over M}\, \delta(\eta-\eta_0) \left[ \xi - {1\over 6} \right] R(\eta) \phi\psi^\pm 
\nonumber \\
&&\qquad\qquad 
+ {1\over 4} {z_4\over M}\, \delta(\eta-\eta_0) \phi^3\psi^\pm 
\biggr\} , \qquad
\label{HintUV}
\end{eqnarray}
which can also be regarded as adding the following boundary action to the theory,
\begin{eqnarray}
S_{\eta=\eta_0} &=& - {1\over 2M}  
\int d^3\vec y\, \sqrt{-h} \biggl\{ 
z_2 \nabla_n^2(\phi\psi^\pm) 
+ {\textstyle{5\over 3}} z_2 K \nabla_n(\phi\psi^\pm) 
\nonumber \\
&&\qquad\qquad\qquad\quad 
+ {\textstyle{2\over 3}} z_2 (\nabla_n K) \phi\psi^\pm 
+ {\textstyle{2\over 3}} z_2 K^2 \phi\psi^\pm
\nonumber \\
&&\qquad\qquad\qquad\quad 
+ z_3 \left[ \xi - {\textstyle{1\over 6}} \right] R \phi\psi^\pm 
+ {\textstyle{1\over 2}} z_4 \phi^3\psi^\pm 
\biggr\} . 
\nonumber \\
&&
\label{actionUV}
\end{eqnarray}
From either perspective, the contribution from these boundary counterterms to the tadpole calculation is 
\begin{eqnarray}
&&\!\!\!\!\!\!\!\!\!\!\!\!\!\!\!\!\!\!\!
\langle \alpha_k^R(\eta_f) | \psi^+_R(x) | \alpha_k^R(\eta_f) \rangle 
\nonumber \\
&=& - {1\over 2} {z_2\over M} {1\over a(\eta_0)}
\partial^2_\eta \left[ a^2(\eta) {\cal G}(\eta_f,\eta) \phi(\eta) \right]_{\eta=\eta_0}
\nonumber \\
&& 
- {1\over 2} {z_3\over M} 
a^3(\eta_0) {\cal G}(\eta_f,\eta_0) \Bigl[ \xi - {1\over 6} \Bigr] \phi(\eta_0) R(\eta_0)
\nonumber \\
&&
- {1\over 4} {z_4\over M} 
a^3(\eta_0) {\cal G}(\eta_f,\eta_0) \phi^3(\eta_0) 
\nonumber \\
&&
+ \cdots . 
\label{leadbndoneUVct}
\end{eqnarray}
As before, the coefficients of the counterterms can be broken into a scale-independent infinite part and a scale-dependent finite part, 
\begin{equation}
z_2 = z_2^\epsilon + \hat z_2(\mu) , 
\quad
z_3 = z_3^\epsilon + \hat z_3(\mu) , 
\quad
z_4 = z_4^\epsilon + \hat z_4(\mu) , 
\label{UVinffin}
\end{equation}
where the infinite part is completely fixed by a renormalization scheme such as the $\overline{\rm MS}$ scheme,
\begin{eqnarray}
z_2^\epsilon &=& {\lambda c_1\over 64i\pi^2} 
\left[ {1\over\epsilon} - \gamma + \ln 4\pi \right] 
\nonumber \\
z_3^\epsilon &=& {\lambda c_1\over 32i\pi^2} 
\left[ {1\over\epsilon} + 1 - \gamma + \ln 4\pi \right] 
\nonumber \\
z_4^\epsilon &=& {\lambda^2 c_1\over 32i\pi^2} 
\left[ {1\over\epsilon} + 1 - \gamma + \ln 4\pi \right]  , 
\label{UVMSbar}
\end{eqnarray}
while the finite part is determined by the Callan-Symanzik equation, 
\begin{eqnarray}
0 &=&
\biggl[ 
\mu {\partial\over\partial\mu} 
+ \beta(\lambda_R) {\partial\over\partial\lambda_R} 
+ m_R \gamma_m(\lambda_R) {\partial\over\partial m_R} 
+ \beta_\xi(\lambda_R) {\partial\over\partial\xi_R} 
\nonumber \\
&&
+ \gamma(\lambda_R)
+ \sum_{n=0}^\infty \hat\beta_n(\lambda_R) {\partial\over\partial\hat z_n}
\biggr] \langle\alpha_k^R(\eta_f) | \psi^+_R(x) | \alpha_k^R(\eta_f)\rangle , 
\nonumber \\
&&
\label{callanbulkNR}
\end{eqnarray}
which has been finally expressed in its complete form.  The sum has been written with an infinite number of boundary $\beta$-functions to represent the possibility of including higher order nonrenormalizable initial conditions.

Adding together the boundary effects in Eq.~(\ref{leadbndoneADUV}) and the counterterms contribution of Eq.~(\ref{leadbndoneUVct}), after applying the $\overline{\rm MS}$ scheme, yields all of the renormalization scale dependence that is associated with the initial state, 
\begin{widetext}
\begin{eqnarray}
\langle \alpha_k^R(\eta_f) | \psi^+_R(x) | \alpha_k^R(\eta_f) \rangle 
&=& - {1\over 2} {1\over M} \biggl[
\hat z_2(\mu) - {\lambda_R c_1\over 32i\pi^2} 
\ln{\mu\over a(\eta_0){\cal M}_R(\eta_0)} 
\biggr]
{1\over a(\eta_0)} \partial^2_\eta \left[ a^2(\eta) {\cal G}(\eta_f,\eta) \phi(\eta) \right]_{\eta=\eta_0}
\nonumber \\
&& 
- {1\over 2} {1\over M} \biggl[
\hat z_3(\mu) - {\lambda_R c_1\over 16i\pi^2} 
\ln{\mu\over a(\eta_0){\cal M}_R(\eta_0)} 
\biggr]
a^3(\eta_0) {\cal G}(\eta_f,\eta_0) \left[ \xi - {1\over 6} \right] \phi(\eta_0) R(\eta_0)
\nonumber \\
&&
- {1\over 4} {1\over M} \biggl[ 
\hat z_4(\mu) - {\lambda^2_R c_1\over 16i\pi^2} 
\ln{\mu\over a(\eta_0){\cal M}_R(\eta_0)} 
\biggr]
a^3(\eta_0) {\cal G}(\eta_f,\eta_0) \phi^3(\eta_0) 
+ \cdots , 
\label{renUV}
\end{eqnarray}
\end{widetext}
which, from the Callan-Symanzik equation, implies that the boundary parameters run as follows,
\begin{eqnarray}
\hat\beta_2(\lambda_R) 
&=& {\lambda_R c_1\over 32i\pi^2} + \cdots 
\nonumber \\
\hat\beta_3(\lambda_R) 
&=& {\lambda_R c_1\over 16i\pi^2} + \cdots 
\nonumber \\
\hat\beta_4(\lambda_R) 
&=& {\lambda_R c_1\over 16i\pi^2} + \cdots . 
\label{NRrunnings}
\end{eqnarray}

So far we have let the parameter $c_1$ be complex, but if the evolution is to remain unitary, it is clear that $c_1$ should be purely imaginary so that its contribution to the boundary counterterm action is real.  By similar reasoning, $d_1$ should also be purely imaginary while $d_0$ is real.

\section{Conclusions}
\label{conclude}

Although the effective theory principle has been widely and advantageously applied in field theory, its role has usually been reserved for describing the evolution of a system from one state to another rather than for describing properties of the actual states involved.  For a field theory in Minkowski space, such a simplification is usually sufficient since scales---including that which divides long-distances from the regime where new physics could modify the structure of the true vacuum state away from a vacuum based on extrapolating the eigenstates of the low energy theory---are time-independent.  Furthermore, the initial and final states are measured in a rather subdued environment, in an asymptotic past or future where the fields no longer interact with each other and where the states become essentially those of the free Minkowski space Hamiltonian.

The conditions prevailing during inflation are dramatically different and can imply a more significant role for the short-distance properties of the states.  In particular, the standard vacuum choice is defined to be that state which resembles the Minkowski vacuum at distances where the curvature is not noticeable, $k\gg H$. However, when the Hubble scale $H$ is itself an appreciable fraction of the scale $M$ for the trans-Planckian physics, we are defining the vacuum precisely in a regime near where the low energy free theory is no longer applicable.  An effective description attempts to provide a generic parameterization of this ambiguity between an extrapolated free vacuum and the true vacuum, without making any assumptions about the physics in the trans-Planckian regime.

Because of the time-evolution of the background, it is not appropriate to choose states based upon their properties in an asymptotically distant past.  Instead, the state is fixed at an ``initial'' time during the regime when the effective theory is predictive, that is, when the perturbative corrections to a general process are still small.  In essence, an effective theory is an application of the principle of decoupling---that the physics of large scales should be relatively independent of the physics at short distances.  To make this idea more precise, we describe the initial state through a power series, as in Eq.~(\ref{ealphser}), which contains terms which either diminish or grow at short distances.  The former are the analogues of the renormalizable part of a bulk effective field theory and they are fixed by the observed long-distance structure of the state.  But it is the latter that contain the signals of trans-Planckian physics and they form the initial-state analogue of the nonrenormalizable operators of a bulk effective theory.  It is important to remember that in this setting the detailed structure of the state is not meant to be part of a complete theory, but only an effective one applicable up to the scale of the unknown trans-Planckian physics.  

The main idea of this article has been to establish the renormalizability of this approach, showing explicitly that once the divergences have been all removed, the signals of the trans-Planckian physics are suppressed by powers of the small ratio of the Hubble scale during inflation and the scale of the new physics, $H/M$.  These new divergences result from summing over the short-distance structure of the initial state.  Thus they appear in a perturbative correction only when we simultaneously sum over arbitrarily high spatial momenta and we evaluate it at the initial time, and as a consequence the counterterms are local operators confined to the same initial surface at which the state was defined.  The power counting for these counterterms parallels that of an ordinary bulk theory, with relevant or marginal boundary counterterms removing the short-distance divergences from the renormalizable part of the initial state and with irrelevant boundary counterterms removing the divergences from the trans-Planckian part. 

One reason for showing the renormalizability of a state with a nontrivial trans-Planckian component is to demonstrate that tree-level calculations, such as that of the primordial power spectrum of fluctuations, are perturbatively stable.  But there is a subtle consequence of this result which has a much more direct effect even on a tree-level calculation.  The search for a renormalizable description of a state is equivalent to finding the correct propagator for the short-distance information which is contained in this state that does not lead to uncontrolled divergences---and it is this same propagator that also appears in the tree-level estimate of a process.  For example, in the Bunch-Davies vacuum, the two-point function and the propagator evaluated for equal times are equivalent, but this equivalence no longer holds for a more general initial state.  The extra term, encoding how the information in the initial state propagates forward, influences the precise calculation of both the primordial power spectrum \cite{schalm,ekp2} and the gravitational back-reaction \cite{schalm2,backreact}.  Quantum mechanically, this term represents the interference between the fields being measured and the initial state information.  As a result, the precise calculation of the trans-Planckian signal requires following this quantum mechanical interference as it affects the classical spectrum of primordial fluctuations, as will be done in \cite{twopoint}.

\begin{acknowledgments}

\noindent
This work was supported in part by DOE grant No.~DE-FG03-91-ER40682 and the National Science Foundation grant No.~PHY02-44801.  

\end{acknowledgments}

\appendix

\section{Kernels}
\label{kernels}

In Sec.~\ref{boundary} we introduced a family of kernel functions,
\begin{equation}
K^{(p)}(\eta) \equiv 
\int {d^3\vec k\over (2\pi)^2}\, 
{e^{-2i\int_{\eta_0}^\eta d\eta'\, \Omega_k(\eta')}
\over \Omega_k^{3-p}(\eta) \Omega_k^p(\eta_0)} , 
\label{Kpkernelenc}
\end{equation}
which occur generically in the loop corrections to a process with a nonrenormalizable initial condition.  The parts that depend on an arbitrary time $\eta$ come from the loop propagator and the part evaluated only on the boundary at $\eta_0$ resulted from the form of the power series we used to describe the initial state, as in Eq.~(\ref{ealphser}).  In fact, our choice for this power series was made to obtain a relatively simple form for the loop integrals, such as the kernel above, when evaluated on the initial boundary.  The structure of the kernels is also chosen so that they only diverge logarithmically in $(\eta-\eta_0)$ after we perform the momentum integral.  Thus if a kernel function appears within an $d\eta$-integral whose integrand, apart from the kernel factor, is well behaved, then the result is finite.  The point of introducing these functions is that the loop integrals we encounter can be written in terms of an appropriate number of derivatives of $K^{(p)}(\eta)$ which we proceed to integrate by parts until all the derivatives have been removed from the kernel still occurring with the conformal time integral.  The boundary terms which result from this process that are evaluated at the initial surface isolate all the new divergences associated with having a nonstandard initial condition on the state.  The remaining momentum integrals can be then regularized, for example by extending the number of spatial dimensions to $3-2\epsilon$. 

In this article, we shall not need to consider more than three derivatives of the kernels, 
\begin{widetext}
\begin{eqnarray}
K^{(p)\, \prime}(\eta) 
&=& 
\int {d^3\vec k\over (2\pi)^3}\, 
e^{-2i\int_{\eta_0}^\eta d\eta'\, \Omega_k(\eta')}
\biggl\{
- {2i\over \Omega_k^{2-p}(\eta) \Omega_k^p(\eta_0)}
- {(3-p) \Omega_k^\prime(\eta)\over \Omega_k^{4-p}(\eta) \Omega_k^p(\eta_0)}
\biggr\}
\nonumber \\
K^{(p)\, \prime\prime}(\eta) 
&=& 
\int {d^3\vec k\over (2\pi)^3}\, 
e^{-2i\int_{\eta_0}^\eta d\eta'\, \Omega_k(\eta')}
\biggl\{
- {4\over \Omega_k^{1-p}(\eta) \Omega_k^p(\eta_0)}
+ {2i(5-2p)\Omega_k^\prime(\eta)\over \Omega_k^{3-p}(\eta) \Omega_k^p(\eta_0)}
- {(3-p)\Omega_k^{\prime\prime}\over \Omega_k^{4-p}(\eta) \Omega_k^p(\eta_0)}
+ {(3-p)(4-p)\Omega_k^{\prime\, 2}\over \Omega_k^{5-p}(\eta) \Omega_k^p(\eta_0)}
\biggr\}
\nonumber \\
K^{(p)\, \prime\prime\prime}(\eta) 
&=& 
\int {d^3\vec k\over (2\pi)^3}\, 
e^{-2i\int_{\eta_0}^\eta d\eta'\, \Omega_k(\eta')}
\biggl\{
{8i\over \Omega_k^{-p}(\eta) \Omega_k^p(\eta_0)}
+ {12(2-p)\Omega_k^\prime(\eta)\over \Omega_k^{2-p}(\eta) \Omega_k^p(\eta_0)}
+ {2i(8-3p)\Omega_k^{\prime\prime}(\eta)\over \Omega_k^{3-p}(\eta) \Omega_k^p(\eta_0)}
- {6i(3-p)^2\Omega_k^{\prime\, 2}(\eta)\over \Omega_k^{4-p}(\eta) \Omega_k^p(\eta_0)}
\nonumber \\
&&\qquad\qquad\qquad\qquad\quad
- {(3-p)\Omega_k^{\prime\prime\prime}(\eta)\over \Omega_k^{4-p}(\eta) \Omega_k^p(\eta_0)}
+ {3(3-p)(4-p)\Omega_k^\prime(\eta)\Omega_k^{\prime\prime}(\eta)\over \Omega_k^{5-p}(\eta) \Omega_k^p(\eta_0)}
- {(3-p)(4-p)(5-p)\Omega_k^{\prime\, 3}(\eta)\over \Omega_k^{6-p}(\eta) \Omega_k^p(\eta_0)}
\biggr\} . 
\nonumber \\
&&
\label{kerneldiff}
\end{eqnarray}
\end{widetext}
In the extreme ultraviolet limit, where the leading part of the generalized frequency approaches $\Omega_k(\eta)\to k$, it is only the first term in each of these expressions which contains a short-distance divergence and which is thus important for determining how the initial state should be renormalized.  Moreover, these kernel functions are all finite except at $\eta=\eta_0$ where there is no longer any oscillatory suppression in the terms that are not already manifestly finite.  Because each of the derivatives in Eq.~(\ref{kerneldiff}) will be integrated by parts at least once, the actual divergences we encounter are found by setting $\eta=\eta_0$ in
\begin{eqnarray}
K^{(p)}(\eta_0) 
&=& \int {d^3\vec k\over (2\pi)^3}\, {1\over \Omega_k^3(\eta_0)}
\nonumber \\
K^{(p)\, \prime}(\eta_0) 
&=& - 2i \int {d^3\vec k\over (2\pi)^3}\, {1\over \Omega_k^2(\eta_0)}
+ \cdots
\nonumber \\
K^{(p)\, \prime\prime}(\eta_0) 
&=& - 4 \int {d^3\vec k\over (2\pi)^3}\, {1\over \Omega_k(\eta_0)}
+ \cdots . 
\label{kerneldiv}
\end{eqnarray}
Note that these divergent parts are independent of the index $p$.  The leading behavior of the adiabatic modes for large momenta is approximated by 
\begin{equation}
\Omega_k(\eta_0) \approx \sqrt{k^2 + a^2(\eta_0){\cal M}^2(\eta_0)} , 
\label{adiabatUV}
\end{equation}
as was derived in Eq.~(\ref{bigOmega2AUV}), so that, up to the prefactors, each of the divergent parts of the kernels can be written in the very general form, 
\begin{equation}
I(0,\alpha) = \int {d^3\vec k\over (2\pi)^3}\, 
{1\over [ k^2 + a^2{\cal M}^2]^{\alpha/2}} . 
\label{genkerint}
\end{equation}
Except for the fact that we are integrating over three rather than four dimensions, this general integral is of the usual form encountered and can be regulated by continuing to an arbitrary real number of dimensions, $3\to 3-2\epsilon$, 
\begin{eqnarray}
I(\epsilon,\alpha) 
&=& \int {d^{3-2\epsilon}\vec k\over (2\pi)^{3-2\epsilon}}\, 
{\mu^{2\epsilon}\over [ k^2 + a^2{\cal M}^2]^{\alpha/2}} 
\nonumber \\
&=& {\sqrt{\pi}\over 8\pi^2} 
{\Gamma(\epsilon- {3-\alpha\over 2})\over \Gamma({\alpha\over 2})}
\Bigl[ {4\pi\mu^2\over a^2{\cal M}^2} \Bigr]^\epsilon 
[ a{\cal M} ]^{3-\alpha} . 
\label{genkerintDR}
\end{eqnarray}
Evaluating the infinite and the finite, nonvanishing, terms for the cases $\alpha=\{ 3,1\}$ yields 
\begin{equation}
K^{(p)}(\eta_0) = {1\over 4\pi^2} 
\left[ {1\over\epsilon} - \gamma 
+ \ln {4\pi\mu^2\over a^2(\eta_0){\cal M}^2(\eta_0)} \right]
+ \cdots 
\label{kpDR} 
\end{equation}
and
\begin{eqnarray}
&&\!\!\!\!\!\!\!\!\!\!\!\!\!\!\!\!\!\!
K^{(p)\, \prime\prime}(\eta_0) 
\label{ddkpDR} \\
&=& {a^2(\eta_0){\cal M}^2(\eta_0)\over 2\pi^2} 
\left[ {1\over\epsilon} + 1 - \gamma 
+ \ln {4\pi\mu^2\over a^2(\eta_0){\cal M}^2(\eta_0)} \right]
+ \cdots . 
\nonumber
\end{eqnarray}
Note that $K^{(p)\, \prime}(\eta_0)$ is finite once its momentum integral has been dimensionally regularized.


\begin{thebibliography}{99}


\bibitem{greens}
H.~Collins and R.~Holman,
Phys.\ Rev.\ D {\bf 71}, 085009 (2005) [hep-th/0501158].

\bibitem{textbooks}
A.~Linde, {\it Particle Physics and Inflationary Cosmology\/} (Harwood Academic, Chur, Switzerland, 1990) [hep-th/0503203]; 
E.~W.~Kolb and M.~S.~Turner, {\it The Early Universe\/} (Addison-Wesley, Reading, MA, 1990); 
A.~Liddle and D.~Lyth, {\it Cosmological Inflation and Large-Scale Structure\/} (Cambridge University Press, Cambridge, England,
2000).

\bibitem{wmap}
H.~V.~Peiris {\it et al.},
Astrophys.\ J.\ Suppl.\  {\bf 148}, 213 (2003) [astro-ph/0302225].

\bibitem{brandenberger}
J.~Martin and R.~H.~Brandenberger,
Phys.\ Rev.\ D {\bf 63}, 123501 (2001) [hep-th/0005209]; 
%
R.~H.~Brandenberger and J.~Martin,
Mod.\ Phys.\ Lett.\ A {\bf 16}, 999 (2001) [astro-ph/0005432].

\bibitem{gary}
R.~Easther, B.~R.~Greene, W.~H.~Kinney and G.~Shiu,
Phys.\ Rev.\ D {\bf 64}, 103502 (2001); 
%
R.~Easther, B.~R.~Greene, W.~H.~Kinney and G.~Shiu,
Phys.\ Rev.\ D {\bf 67}, 063508 (2003); 
%
G.~Shiu and I.~Wasserman,
Phys.\ Lett.\ B {\bf 536}, 1 (2002); 
%
R.~Easther, B.~R.~Greene, W.~H.~Kinney and G.~Shiu,
Phys.\ Rev.\ D {\bf 66}, 023518 (2002).

\bibitem{transplanck}
J.~C.~Niemeyer,
Phys.\ Rev.\ D {\bf 63}, 123502 (2001); 
%
A.~Kempf,
Phys.\ Rev.\ D {\bf 63}, 083514 (2001); 
%
J.~C.~Niemeyer and R.~Parentani,
Phys.\ Rev.\ D {\bf 64}, 101301 (2001); 
%
A.~Kempf and J.~C.~Niemeyer,
Phys.\ Rev.\ D {\bf 64}, 103501 (2001); 
%
A.~A.~Starobinsky,
Pisma Zh.\ Eksp.\ Teor.\ Fiz.\  {\bf 73}, 415 (2001)
[JETP Lett.\  {\bf 73}, 371 (2001)]; 
%
L.~Hui and W.~H.~Kinney,
Phys.\ Rev.\ D {\bf 65}, 103507 (2002); 
%
S.~Shankaranarayanan,
Class.\ Quant.\ Grav.\  {\bf 20}, 75 (2003); 
%
S.~F.~Hassan and M.~S.~Sloth,
Nucl.\ Phys.\ B {\bf 674}, 434 (2003); 
%
K.~Goldstein and D.~A.~Lowe,
Phys.\ Rev.\ D {\bf 67}, 063502 (2003); 
%
V.~Bozza, M.~Giovannini and G.~Veneziano,
JCAP {\bf 0305}, 001 (2003); 
%
G.~L.~Alberghi, R.~Casadio and A.~Tronconi,
Phys.\ Lett.\ B {\bf 579}, 1 (2004); 
%
J.~Martin and R.~Brandenberger,
Phys.\ Rev.\ D {\bf 68}, 063513 (2003); 
%
U.~H.~Danielsson,
Phys.\ Rev.\ D {\bf 66}, 023511 (2002); 
%
U.~H.~Danielsson,
JHEP {\bf 0207}, 040 (2002); 
%
R.~H.~Brandenberger and J.~Martin,
Int.\ J.\ Mod.\ Phys.\ A {\bf 17}, 3663 (2002).

\bibitem{cliff}
C.~P.~Burgess, J.~M.~Cline, F.~Lemieux and R.~Holman,
JHEP {\bf 0302}, 048 (2003) [hep-th/0210233]; 
%
C.~P.~Burgess, J.~M.~Cline and R.~Holman,
JCAP {\bf 0310}, 004 (2003) [hep-th/0306079]; 
%
C.~P.~Burgess, J.~Cline, F.~Lemieux and R.~Holman,
astro-ph/0306236.

\bibitem{fate}
H.~Collins, R.~Holman and M.~R.~Martin,
Phys.\ Rev.\ D {\bf 68}, 124012 (2003) [hep-th/0306028]; 
%
H.~Collins and M.~R.~Martin,
Phys.\ Rev.\ D {\bf 70}, 084021 (2004) [hep-ph/0309265]; 
%
H.~Collins,
hep-th/0312144.

\bibitem{ekp1}
R.~Easther, W.~H.~Kinney and H.~Peiris,
JCAP {\bf 0505}, 009 (2005) [astro-ph/0412613].

\bibitem{kaloper}
N.~Kaloper, M.~Kleban, A.~E.~Lawrence and S.~Shenker,
Phys.\ Rev.\ D {\bf 66}, 123510 (2002); 
%
N.~Kaloper, M.~Kleban, A.~Lawrence, S.~Shenker and L.~Susskind,
JHEP {\bf 0211}, 037 (2002).

\bibitem{schalm}
K.~Schalm, G.~Shiu and J.~P.~van der Schaar,
JHEP {\bf 0404}, 076 (2004) [hep-th/0401164]; 
%
B.~Greene, K.~Schalm, J.~P.~van der Schaar and G.~Shiu,
eConf {\bf C041213}, 0001 (2004) [astro-ph/0503458].

\bibitem{schalm2}
B.~R.~Greene, K.~Schalm, G.~Shiu and J.~P.~van der Schaar,
JCAP {\bf 0502}, 001 (2005) [hep-th/0411217];
%
K.~Schalm, G.~Shiu and J.~P.~van der Schaar,
AIP Conf.\ Proc.\  {\bf 743}, 362 (2005) [hep-th/0412288].

\bibitem{ekp2}
R.~Easther, W.~H.~Kinney and H.~Peiris,
astro-ph/0505426.

\bibitem{emil}
P.~R.~Anderson, C.~Molina-Paris and E.~Mottola,
hep-th/0504134.

\bibitem{planck}
J.~A.~Tauber on behalf of ESA and the Planck Scientific Collaboration, 
Adv.\ Space Res.\ {\bf 34}, 491 (2004).

\bibitem{ska}
Two such examples include the Square Kilometre Array and the Cosmic Inflation Probe:  respectively, 
S.~Rawlings, F.~B.~Abdalla, S.~L.~Bridle, C.~A.~Blake, C.~M.~Baugh, L.~J.~Greenhill and J.~M.~van der Hulst,
New Astron.\ Rev.\  {\bf 48}, 1013 (2004) [astro-ph/0409479] 
and  
G.~J.~Melnick, G.~G.~Fazio, V.~Tolls, D.~T.~Jaffe, K.~Gebhardt, V.~Bromm, E.~Komatsu, and R.~A.~Woodruff,
Am.\ Astron.\ Soc.\ Meeting {\bf 205} (2004) 10006.

\bibitem{eft}
H.~Georgi, ``Weak Interactions And Modern Particle Theory,'' (Benjamin/Cummings, Menlo Park, California, 1984); 
%
H.~Georgi, ``Effective field theory,''
Ann.\ Rev.\ Nucl.\ Part.\ Sci.\  {\bf 43}, 209 (1993); 
%
I.~Z.~Rothstein,
``TASI lectures on effective field theories,''
hep-ph/0308266.

\bibitem{witten}
In de Sitter space, for example, refer to E.~Witten,
hep-th/0106109.

\bibitem{schwinger}
J.~S.~Schwinger,
J.\ Math.\ Phys.\  {\bf 2}, 407 (1961).

\bibitem{keldysh}
L.~V.~Keldysh,
Zh.\ Eksp.\ Teor.\ Fiz.\  {\bf 47}, 1515 (1964)
[Sov.\ Phys.\ JETP {\bf 20}, 1018 (1965)].

\bibitem{kt}
K.~T.~Mahanthappa, Phys.\ Rev. {\bf 126}, 329 (1962); P.~M.~Bakshi and K.~T.~Mahanthappa, J.\ Math.\ Phys.\ {\bf 41}, 12 (1963).

\bibitem{weinberginin}
S.~Weinberg,
hep-th/0506236.

\bibitem{twopoint}
H.~Collins and R.~Holman, {\it in progress\/}.

\bibitem{mcnees}
F.~Larsen and R.~McNees,
JHEP {\bf 0307}, 051 (2003) [hep-th/0307026]; 
F.~Larsen and R.~McNees,
JHEP {\bf 0407}, 062 (2004) [hep-th/0402050].

\bibitem{bunch}
T.~S.~Bunch and P.~C.~Davies,
Proc.\ Roy.\ Soc.\ Lond.\ A {\bf 360}, 117 (1978).

\bibitem{einhorn}
M.~B.~Einhorn and F.~Larsen,
Phys.\ Rev.\ D {\bf 67}, 024001 (2003) [hep-th/0209159]; 
%
M.~B.~Einhorn and F.~Larsen,
Phys.\ Rev.\ D {\bf 68}, 064002 (2003) [hep-th/0305056].

\bibitem{banks}
T.~Banks and L.~Mannelli,
Phys.\ Rev.\ D {\bf 67}, 065009 (2003)
[hep-th/0209113].

\bibitem{lowe}
K.~Goldstein and D.~A.~Lowe,
Phys.\ Rev.\ D {\bf 69}, 023507 (2004) [hep-th/0308135].

\bibitem{taming}
H.~Collins and R.~Holman,
Phys.\ Rev.\ D {\bf 70}, 084019 (2004) [hep-th/0312143]; 
%
H.~Collins,
Phys.\ Rev.\ D {\bf 71}, 024002 (2005) [hep-th/0410229]; 
%
H.~Collins,
hep-th/0410228.

\bibitem{alpha}
N.~A.~Chernikov and E.~A.~Tagirov,
Annales Poincare Phys.\ Theor.\ A {\bf 9}, 109 (1968); 
E.~A.~Tagirov,
Annals Phys.\  {\bf 76}, 561 (1973); 
%
E.~Mottola,
Phys.\ Rev.\ D {\bf 31}, 754 (1985); 
%
B.~Allen,
Phys.\ Rev.\ D {\bf 32}, 3136 (1985).

\bibitem{neft}
D.~Boyanovsky, H.~J.~de Vega, R.~Holman, D.~S.~Lee and A.~Singh,
Phys.\ Rev.\ D {\bf 51}, 4419 (1995) [hep-ph/9408214].

\bibitem{weinberg}
S.~Weinberg,
Phys.\ Rev.\ D {\bf 9}, 3357 (1974).

\bibitem{wilson}
K.~G.~Wilson and J.~B.~Kogut,
Phys.\ Rept.\  {\bf 12}, 75 (1974).

\bibitem{backreact}
M.~Porrati,
Phys.\ Lett.\ B {\bf 596}, 306 (2004) [hep-th/0402038].
%
M.~Porrati,
hep-th/0409210.
%
F.~Nitti, M.~Porrati and J.~W.~Rombouts,
hep-th/0503247.

\end{thebibliography}
\end{document}